\documentclass[sigconf,authorversion]{acmart}
\AtBeginDocument{%
  \providecommand\BibTeX{{%
    \normalfont B\kern-0.5em{\scshape i\kern-0.25em b}\kern-0.8em\TeX}}}

\usepackage{subfiles}
\usepackage{caption}
\usepackage{subcaption}
\usepackage{multirow}
\usepackage[compress]{cleveref}
\usepackage{enumitem}

\crefname{subfigure}{subfigure}{subfigures}
\Crefname{subfigure}{Subfigure}{Figures}

\crefrangelabelformat{Figures}{(#3#1#4--#5\crefstripprefix{#1}{#2}#6)}

\DeclareMathOperator*{\argmax}{argmax}
\newcommand{\B}{\textbf}

\copyrightyear{2022}
\acmYear{2022}
\setcopyright{rightsretained}

\acmConference[KDD '22]{Proceedings of the 28th ACM SIGKDD Conference on Knowledge Discovery and Data Mining}{August 14--18, 2022}{Washington, DC, USA}
\acmBooktitle{Proceedings of the 28th ACM SIGKDD Conference on Knowledge Discovery and Data Mining (KDD '22), August 14--18, 2022, Washington, DC, USA}
\acmISBN{978-1-4503-9385-0/22/08}
\acmDOI{10.1145/3534678.3539432}

\usepackage{etoolbox}
\makeatletter
\patchcmd{\maketitle}{\@copyrightpermission}{
   \begin{minipage}{0.3\columnwidth}
     \href{https://creativecommons.org/licenses/by/4.0/}{\includegraphics[width=0.90\textwidth]{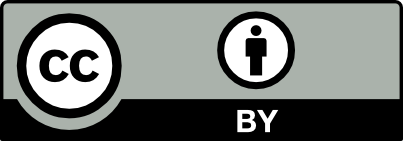}}
   \end{minipage}\hfill
   \begin{minipage}{0.7\columnwidth}
     \href{https://creativecommons.org/licenses/by/4.0/}{This work is licensed under a Creative Commons Attribution International 4.0 License.}
   \end{minipage}

   \vspace{5pt}
}{}{}

\begin{document}

\title{PARSRec: Explainable Personalized Attention-fused Recurrent Sequential Recommendation Using Session Partial Actions}

\author{Ehsan Gholami}
\email{egholami@ucdavis.com}
\orcid{0000-0001-8676-0497}
\affiliation{%
  \institution{UC Davis}
  \streetaddress{1 Shields Ave.}
  \city{Davis}
  \state{California}
  \country{USA}
  \postcode{95616}
}

\author{Mohammad Motamedi}
\email{mmotamedi@ucdavis.edu}
\orcid{0000-0003-0120-8738}
\affiliation{%
  \institution{UC Davis}
  \streetaddress{1 Shields Ave.}
  \city{Davis}
  \state{California}
  \country{USA}
  \postcode{95616}
}

\author{Ashwin Aravindakshan}
\email{aaravind@ucdavis.com}
\orcid{0000-0002-5609-9746}
\affiliation{%
  \institution{Graduate School of Management, UC Davis}
  \streetaddress{1 Shields Ave.}
  \city{Davis}
  \state{California}
  \country{USA}
  \postcode{95616}
}

\begin{abstract}
The emerging meta- and multi-verse landscape is yet another step towards the more prevalent use of already ubiquitous online markets. In such markets, recommender systems play critical roles by offering items of interest to the users, thereby narrowing down a vast search space that comprises hundreds of thousands of products. Recommender systems are usually designed to learn common user behaviors and rely on them for inference. This approach, while effective, is oblivious to subtle idiosyncrasies that differentiate humans from each other. Focusing on this observation, we propose an architecture that relies on common patterns as well as individual behaviors to tailor its recommendations for each person. Simulations under a controlled environment show that our proposed model learns interpretable personalized user behaviors. Our empirical results on Nielsen Consumer Panel dataset indicate that the proposed approach achieves up to 27.9\% performance improvement compared to the state-of-the-art.
\end{abstract}



\maketitle
\pagestyle{plain}

\section{INTRODUCTION}

The task of recommender systems is to delineate users' interests accurately. Recommender systems help providers offer viable alternatives to users as they navigate amongst a vast number of available choices. They achieve this by leveraging users' historical behavior to extract meaningful patterns that help predict users' future interests. These patterns often change over time and are heterogeneous across users. Therefore, deriving functional patterns becomes increasingly challenging with growing numbers of users, items, and user-item actions. Recommender systems focus on capturing these evolving, diverse, and high-dimensional behaviors.
\begin{figure}
\centering
     \includegraphics[width=\columnwidth]{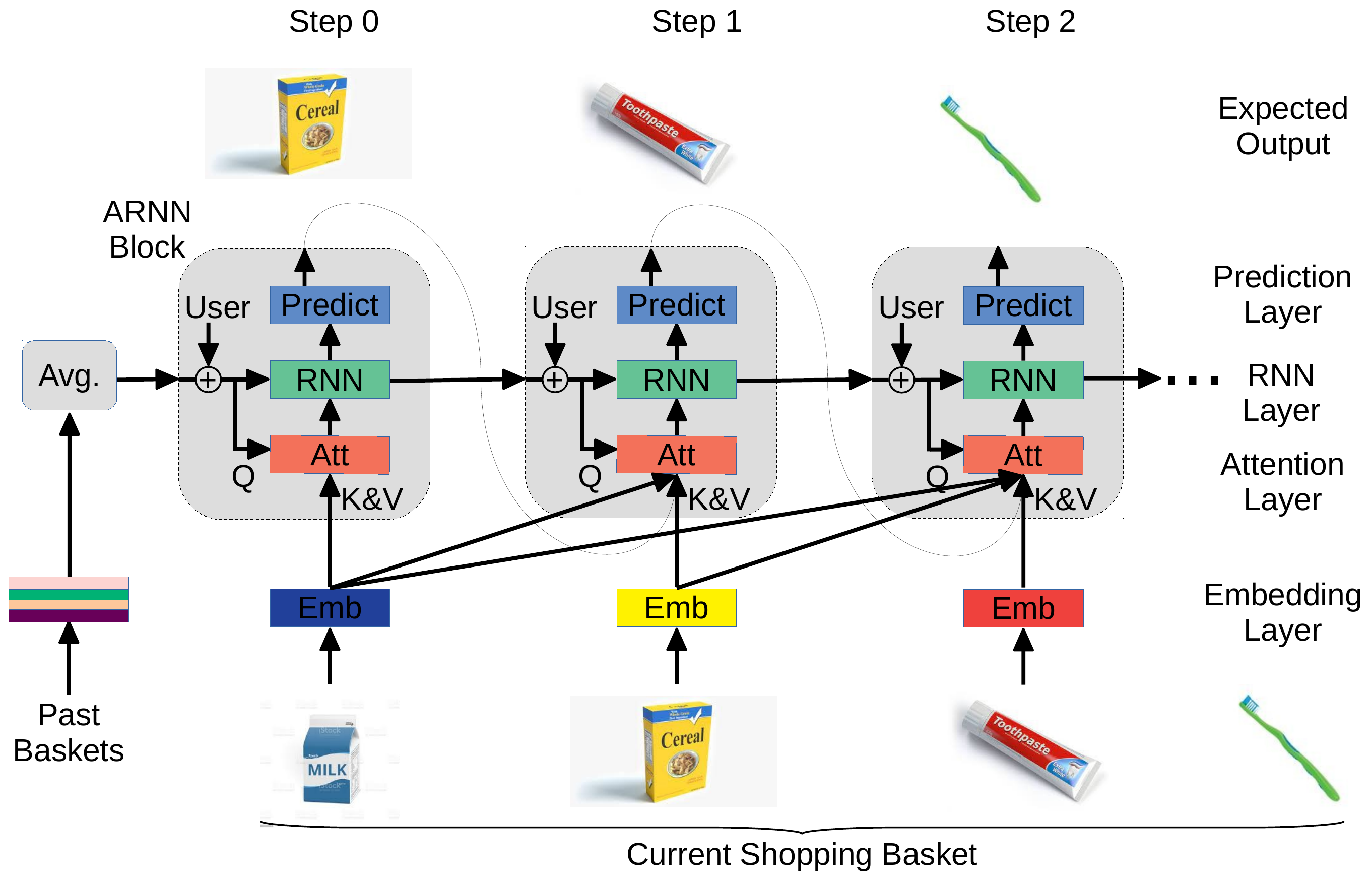}
     \caption{Coarse Architecture of PARSRec: In every step, first, the Attention module investigates the items that currently exist in the shopping basket to identify those items that will considerably impact the selection of the next item. Then, it represents such items in the latent space as a hint to the recurrent network. Furthermore, the user-aware nature of the proposed architecture makes it possible to leverage users' idiosyncrasies for predicting and suggesting the next item.}
     \label{fig:arch}
     \Description{Architecture of PARSRec model. A modified RNN that utilizes attention in it.} 
\end{figure}

Two types of recommender systems have gained popularity in recent research,~i)~sequential and~ii)~session-based recommenders. Sequential recommenders
often consider all historical user actions as a single ordered sequence
and try to successively infer each user action based on the user’s prior
actions in the sequence. Session-based recommenders only leverage the user's most recent actions called session (e.g., anonymous online shopper without an existing historical behavior). The state-of-the-art approaches benefit from deep neural nets to enhance the performance of recommendation tasks. Recurrent Neural Networks (RNN) \cite{donkers2017sequential,quadrana2017personalizing} and their improved variants such as Gated Recurrent Units (GRU) \cite{hidasi2015session,hidasi2018recurrent} and Long Short-Term Memory (LSTM) \cite{feng2019deep} often capture all the previous actions of the user in the past via hidden states. This strategy allows them to understand complex user behaviors. However, RNNs suffer from long-range dependencies because long-range back-propagated gradients can vanish or explode. LSTM and GRU prove effective in some fields by resolving this issue but have their limitations in the field of sequential recommendation. They tend to summarize the session information into a single representation. For example, in machine translation, words within a sentence are related to each other (in various degrees). However, not all items within a shopping basket are necessarily related. For example, in the shopping basket \emph{(milk, cereal, laundry detergent)}, the choice of laundry detergent could be completely independent of milk and cereal. Encoding the entire session into one (or a few) representation(s) will entangle irrelevant information together and would possibly make it harder for the decoder to detangle them. These methods assume a natural order to historical user actions, which does not always hold in real-world applications. Moreover, they tend to become slow when processing long sequences due to their sequential nature.

More recent recommender systems use attention mechanisms to overcome the issues mentioned above by identifying a smaller number of actions most relevant to the next item recommendation. They are effective in many applications and can provide helpful interpretable visuals for item-item relationships. The self-attention mechanism in Transformers learns long-range global item-item relationships and utilizes that along with the items in the user's historical actions to learn heterogeneous user behaviors implicitly. They require less dense data and can run faster in parallel. However, the state-of-the-art transformer recommender systems usually can handle limited input length, limiting the number of user actions\cite{sun2019bert4rec, kang2018self, chen2019behavior}. This is due to the extra positional embedding required to capture the relative or absolute order of the items in a sequence. On top of that, item relations generally differ from one user to another. For instance, user $A$ might purchase milk for cookies, while user $B$ might purchase milk for cereal and independently buy cookies simultaneously. Most attention-based models capture only the universal item-item relationships based only on item sequence and lack personalized interpretable user behavior representations.

Inspired by these methods, we propose PARSRec, a \textbf{P}ersonalized \textbf{A}ttention-based \textbf{R}ecurrent \textbf{S}equential \textbf{Rec}ommender that fuses the attention mechanism with RNNs, illustrated in Figure \ref{fig:arch}, to address the limitations mentioned above. Our framework partitions a user's actions into two groups: \emph{i)} information from past sessions and \emph{ii)} items interacted with so far during the current session. There are multiple reasons for this approach. In a wide range of applications, the partial knowledge of the current session provides much richer information than the previous sessions \cite{he2017translation}. For example, one may purchase cereal and milk on a trip to a grocery store. Knowing the partial information of the cereal is more likely to help predict the next item, milk, accurately compared to the shopping baskets in the past. In another example, the list of the music tracks that a user listens to in a session is more likely to be related to other songs in the same session than tracks in other sessions \cite{hansen2020contextual}. Second, in many real-world cases, there is no order to the items in a session (e.g., items within a basket), whereas the sessions themselves may follow a chronological order. Lastly, our model eliminates the need for positional embeddings by such partitioning of the historical interactions. This, in turn, reduces the network's memory footprint.

The hidden states of the RNN network in our model carry the user information and the user's historical behavior. The attention layer, which is agnostic to orders, uses the hidden state to determine which items within the current session are more relevant in predicting the next item. PARSRec outperforms the state-of-the-art methods on Nielsen's real-world Consumer Panel dataset, as detailed in Section \ref{sec:experiments}. The model extracts interpretable personalized user behavior by using explicit user representations in the attention layer's queries. We show that PARSRec can accurately explain personalized user behavior in a controlled environment on a synthetic dataset. This powerful explanation allows the provider to fully understand the underlying user behavior beyond a simple next item recommendation to make informed decisions on many tasks. Examples of tasks that can benefit from this knowledge are assortment optimization, assortment allocation (what items go together on the same shelf or a webpage design), and personalized coupons, discounts, and displays. Our recurrent model capacity is independent of the sequence length. Its complexity depends only on the number of items within a session which is usually small as detailed in Section \ref{sec:experiments}. The network can utilize any length of user history without increasing the capacity or complexity of the model. The key contributions of our work are:\\
\begin{itemize}[noitemsep,topsep=0pt,leftmargin=*]
\item we propose a model that uses attention layers combined with RNNs for the task of sequential recommendation. We show that our model outperforms various state-of-the-art methods on synthetic and real-world data under different evaluation metrics.
\item we test our model in a controlled environment of a synthetic dataset. we show that our model learns personalized user behaviors and offers interpretable results through visualizing item relationships.
\item we conduct an ablation study on variations of the proposed model to evaluate the contribution of components of the model and report the most effective architecture.
\end{itemize}

\section{RELATED WORK}
We review the works on closely related recommender systems to our framework.

\textbf{General Recommendation}:
Collaborative Filtering (CF) is one of the classic approaches in the field of recommender systems \cite{koren2015advances,ricci2011introduction,hidasi2019cutting}. CF infers user preferences from their historical interactions. Matrix Factorization (MF) is a successful CF method that uses a shared space to represent both users and items \cite{koren2009matrix,mnih2007probabilistic}. More recent approaches use deep learning to improve the effectiveness of models. However, the state-of-the-art works rely on deep learning to provide recommendations \cite{yu2021visually,guo2020learning,hansen2020contextual,naumov2019deep,tang2018personalized}.

\textbf{Sequential Recommendation}:
Studies of sequential recommendation aim to extract item transitions in a sequence of items a user interacts with. Markov Chain (MC) models capture such transitions. Factorizing Personalized Markov Chain (FPMC) \cite{rendle2010factorizing} and its extension Hierarchical Representation Model (HRM) \cite{wang2015learning} combine MF and MC to extract a personalized item transition. Recurrent models have also shown promising results in the field of sequential recommendation. RNNs and their variants (e.g., GRU and LSTM) have been used for modeling user interaction sequences. Most RNN models encode a user's historical behavior into a representation vector and use that along with the current interaction as the input to the model to predict the next action. DREAM \cite{yu2016dynamic} adopted MCs in a recurrent setting to create dynamic representations of users. GRU4Rec \cite{hidasi2015session} and GRU4Rec$^+$ \cite{hidasi2018recurrent} use Gated Recurrent Units for session-based recommendations. Some works utilize memory networks to store users' actions in RNNs for recommendation \cite{chen2018sequential,huang2018improving}. BINN \cite{li2018learning} uses two components to capture long and short-term preferences in the RNN setting.

\textbf{Attention-based Recommendation}:
Attention mechanism has gained popularity in many fields (e.g., machine translation) due to its promising performance and interpretability. Recent state-of-the-art recommender systems use the attention mechanism. NARM uses attention with encoder to model the user’s sequential behavior and the user’s session purpose \cite{li2017neural}. ACA-GRU leverages attention mechanism to build a context-aware recommender system \cite{yuan2020attention}. STAMP uses attention to capture users’ general interests from the long-term memory of a session context and users’ current interests from the short-term memory \cite{liu2018stamp}. KGAT \cite{wang2019kgat} uses the attention mechanism on knowledge graphs for the recommendation. SASRec \cite{kang2018self} and BERT4Rec \cite{sun2019bert4rec} use uni-directional and bi-directional self-attention mechanisms (i.e., Transformers) to capture item-item relationships and have achieved state-of-the-art performances. TiSASRec \cite{li2020time} incorporates time intervals in attention mechanism for a time-aware recommender system.

Existing methods that use attention mechanisms usually learn an implicit representation of users using the global item-item relationships. These methods often limit the length of the input sequence (number of user's historical actions) and require cropping the input to a pre-set max-length. Some methods also assume a rigid order to the sequence of user actions. We seek to design a model that addresses these limitations. Our model outperforms the state-of-the-art methods and learns explainable personalized item-item relationships that provide insights into user choices.
\section{PROBLEM STATEMENT}

In recommender systems, a set of users, $U = \{u_1, u_2, ..., u_{|U|}\}$, interact with a set of items, $V = \{v_1, v_2, ..., v_{|V|}\}$. Examples of user-item interactions are purchasing an item, listening to a track, watching a movie, or clicking on a link. A sequence of interactions by user $u \in U$, denoted by $S_u$, is partitioned into sessions $(S^u_{t_1}, S^u_{t_2}, ..., S^u_{t_{|S_u|}})$. Each session $S^u_{t_i} \subseteq V$ is the interaction of user $u$ with a set of items $\{v^{(S)}_j|v^{(S)}_j \in V, 1<j<n^u_i\}$ at time $t_i$.
The total number of items in the session is denoted by $n^u_i = |S^u_{t_i}|$, and the sessions are in chronological order ($t_1<t_2<...<t_{|S_u|}$). An example of $S^u_{t_i}$ would be a shopping basket or a session of listening to music tracks. In this paper, we use the terms session and basket interchangeably. We assume there is no specific chronological order to items within a session. For instance, a shopping basket at a brick-and-mortar store does not provide a meaningful order. However, if there is a meaningful order to items within a session, an ideal solution would be able to capture that relationship as well. The task of sequential recommendation is to predict the next item in the session $v^{(S)}_{j+1}$ given the history of user behavior $(S^u_{t_1}, ..., S^u_{t_{i-1}})$, and the subset of items  $[v^{(S)}_1, ..., v^{(S)}_j]$ that user has interacted with so far during the current session $S^u_{t_i}$.
\section{MODEL ARCHITECTURE}

\begin{figure}
    \centering
    \includegraphics[width=0.9\columnwidth]{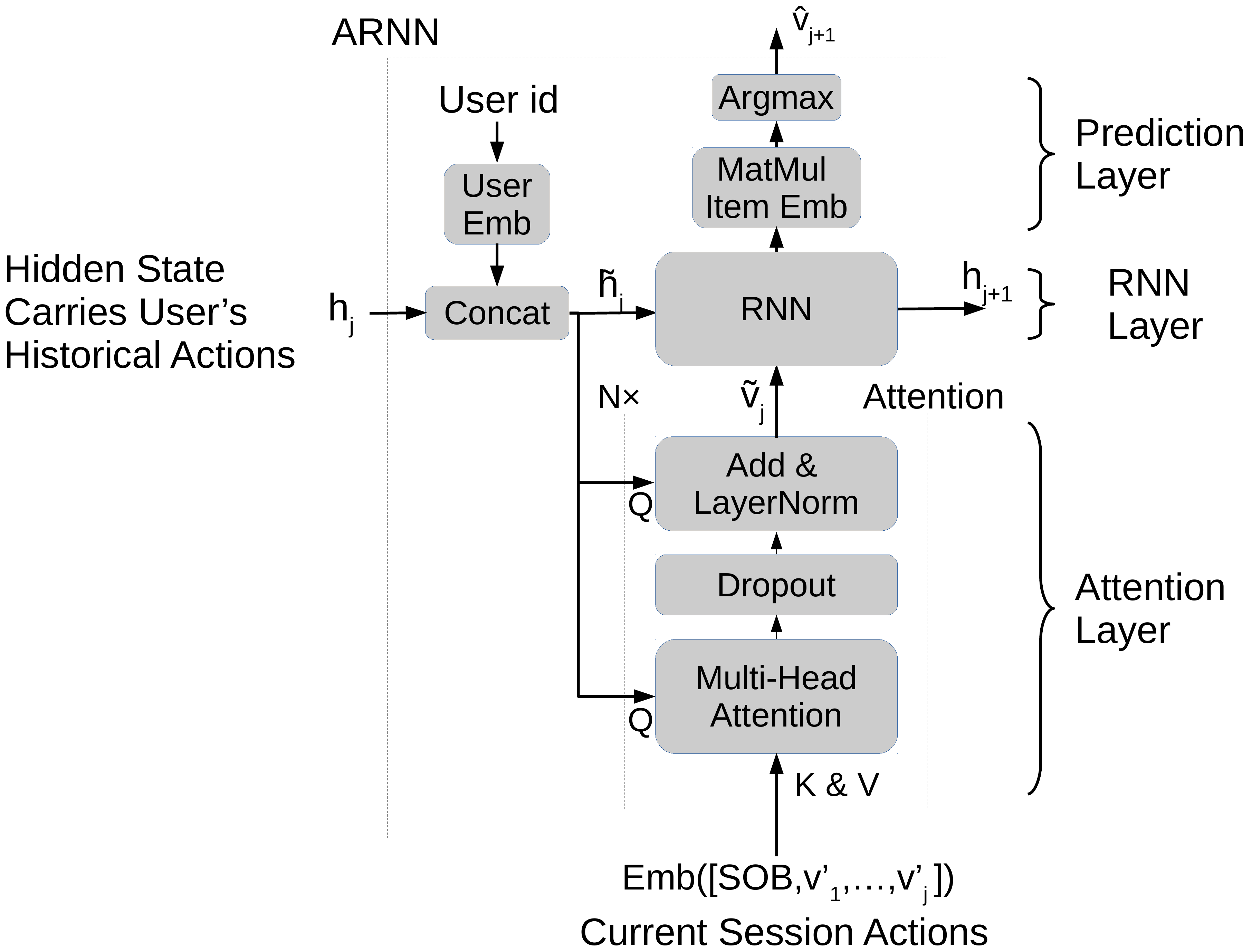}
    \caption{ARNN: Attention-fused RNN Block}
    \label{fig:arnn}
    \Description{Architecture of ARNN}
\end{figure}

In this section, we introduce a new sequential recommendation model called \textbf{PARSRec}. A \textbf{P}ersonalized \textbf{A}ttention-based \textbf{R}ecurrent \textbf{S}equential \textbf{Rec}ommender that combines the power of recurrent neural networks (RNN) and scaled dot-product attention into a modified \textbf{A}ttention-fused \textbf{RNN}: \textbf{ARNN} (Figure \ref{fig:arnn}). The outer structure of the model is similar to vanilla RNN, i.e., each ARNN block receives a hidden state and an input and produces an output and the next hidden state. However, the inner architecture of the ARNN block is different than that of vanilla RNN. All ARNN blocks in Figure \ref{fig:arnn} use the same parameters. Next we detail the architecture.
\subsection{Embedding Layer}
We build two embedding tables, $\textbf{E}^U \in   \mathbb{R} ^ {|U| \times d_u}$ and $\textbf{E}^V \in   \mathbb{R} ^ {|V| \times d_v}$, to represent users and items in the latent space, respectively, explained in detail in Appendix \ref{sec:app-model}. We denote the latent dimensions of users by $d_u$ and items by $d_v$. 

\subsection{Attention Layer}\label{ssec:attention}
The attention layer can learn dependencies between two representations regardless of their position in a sequence. We use the attention layer to identify
the existing items in the current session that will considerably impact the selection of the next merchandise. The attention module in our model consists of three sub-layers:

\textbf{Multi-Head Attention Block:} We adopt the multi-head attention module described in \cite{vaswani2017attention}. It is shown that learning from multiple sub-spaces of representations is more flexible than a single representation. Multi-head attention inherently breaks down the representations into smaller sub-spaces, applies attention on each sub-space, and then concatenates the outputs back into a single representation. The attention block is a scaled dot-product of three vectors, key \textbf{K}, value \textbf{V}, and query \textbf{Q}. Intuitively, attention is a weighted sum of rows in $\B{V}$ and weights are defined by similarity of rows in matrices $\B{Q}$ and $\B{K}$. In our model, $\B{K}$ and $\B{V}$ are the same and their rows are embeddings of items $\{v^{(S)}_1, ..., v^{(S)}_{n^u_i}\}$. The matrix $\B{Q}$, however, is different and is the concatenation of user embedding and the previous hidden state. Similar to \cite{vaswani2017attention}, to alleviate overfitting, help with stability, and speed up the training process, we add Layer Normalization (LN), and Dropout. We refer to the described architecture as attention layer in the rest of the paper.

\textbf{Stacking Attention Layers}: Utilizing more numbers ($N>1$) of attention layers hierarchically allows the model to learn richer and more complex relationships. We investigate this structure by stacking multiple attention layers using the same query \textbf{Q} and feeding the output of the lower layer as the next layer's key \textbf{K} and value \textbf{V}. It may be beneficial to use multiple attention layers depending on the application and complexity of the relationships.

\subsection{Recurrent Architecture} We combine the attention layer with a vanilla RNN by using the attention layer output as the input $\tilde{v}_j$ to the RNN block and attention query \textbf{Q} as the input hidden state $\tilde{h}_j$ to the RNN block.

The hidden state $\tilde{h}_j$ carries the user information and the current state of the session. We concatenate user embedding to the hidden state at every step to ensure it does not vanish as we progress. The initial hidden state encodes the interaction history, which is the weighted average of embeddings of items in the users' historical actions.
The attention layer identifies the items whose presence in the basket is expected to highly impact the selection of the next item. Subsequently, these items are offered to the RNN to enhance its prediction accuracy.

\textbf{Prediction Layer:} We use the output of the RNN block at each step to predict the next item. The output of the RNN layer is multiplied by $\textbf{E}^V$ to provide a similarity vector of size $|V|$. The indices with higher values represent items that are more likely to be interacted with next. We can rank this vector and make recommendations based on that. The objective loss function will convert the output vector to a probability vector using a softmax layer, which is discussed in the next subsection.
\subsection{Loss Function} We adopt cross entropy loss function as our objective function:
\begin{equation}
    - log\Big(\frac{exp(y[target])}{\sum_{k}{exp(y[k])}}\Big)
\end{equation}
\noindent where $y$ is the output of ARNN layer and $target$ is the ground truth item to be predicted $v'_{j+1}$. This objective function combines LogSoftmax and negative log-likelihood loss. The loss is averaged across observations of each mini-batch. We also ignore any $EOB$ (\textbf{E}nd \textbf{O}f \textbf{B}asket) padded items in training.
\subsection{Training}  We train the model to predict the items in the current session $S^u_i$ sequentially given a user and their historical behavior. A basket of size $n^u_i$ will have $n^u_i$ prediction steps. Following a common practice in sequential recommender systems \cite{kang2018self} and machine translation models \cite{vaswani2017attention}, we benefit from teacher enforcing during the training. However, since we do not assume any natural order for items in the session, it is beneficial to know which items in the basket are the most related items to predict the next item $j$. Some recommenders use bidirectional learning that utilizes both left and right items for prediction \cite{sun2019bert4rec}. The idea is to randomly mask an item (or items) and use remaining information on both left and right to predict masked item(s) \cite{baevski2019cloze, devlin2018bert}. To alleviate the same issue, we take a different approach and modify our teacher enforcing at each step as follows:
\begin{equation}
  v'_j =
    \begin{cases}
      SOB & j = 0\\
      \begin{cases}
            \hat{v}_j & if\ \hat{v}_j \in \{v'_{j+1}, ..., v'_{n^u_i}\}\\
            rand(\{v'_{j+1}, ..., v'_{n^u_i}\}) & if\ \hat{v}_j \notin \{v'_{j+1}, ..., v'_{n^u_i}\}\\
      \end{cases}
      & 0<j<n^u_i\\
      EOB & j \ge n^u_i
    \end{cases}       
\end{equation}
\noindent where $SOB$ and $EOB$ are \textbf{S}tart and \textbf{E}nd \textbf{O}f \textbf{B}asket tokens. In other words, we start with $SOB$ as input of step $0$, and if the predicted output in each step is in the rest of the basket, we add that to inputs of the next step. Otherwise, we randomly pick an item from the remaining items in the basket to perform the teacher enforcing. This method of teacher enforcing has the benefit of bringing related items within a session closer to each other in position and the attention layer can capture those relationships. For instance, a basket in the dataset might be \emph{(cereal, toothbrush, milk, toothpaste)}. The aforementioned teacher enforcing method brings \emph{(cereal, milk)} and \emph{(toothbrush, toothpaste)} closer in position by accurately predicting the item-item relationships.

To perform training in mini-batches, we randomly put baskets of similar sizes in the same mini-batch to avoid unnecessary calculation steps. We pad all baskets on the left with $SOB$ and if some baskets within the mini-batch are of different sizes, we pad them with $EOB$ on the right. We use \emph{Adam} and \emph{Sparse Adam} optimizers to optimize network's non-sparse and embedding parameters, respectively. Both optimizers adaptively estimate the moments. We also examined Stochastic Gradient Descent (SGD) optimizer and found similar results.

\textbf{Causality:} To avoid leaking information from the future to the prediction of the current step, we only feed the model with the inputs seen at previous steps. The only added item will be either the prediction of the last step or a randomly selected item from the remaining basket. We do not use any extra information from the remaining of the session (including the item we are predicting at the current step). Our modified teacher enforcement makes sure no future information is leaked to the past.
\section{EXPERIMENTAL RESULTS}\label{sec:experiments}
\subsection{Datasets}

We evaluate our model on two datasets: a controlled synthetic dataset and a real-world dataset. We discuss the details of each data preprocessing next. 
\begin{table}
  \caption{Post preprocessing datasets statistics}
  \label{tab:data_stats}
  \resizebox{\columnwidth}{!}{%
  \begin{tabular}{lrrrrr}
    \toprule
    Dataset & \#users & \#items & \#actions & avg. actions/user & Density\\
    \midrule
    Synthetic & 8,192 & 2,000 & 5.8M & 710 & 0.10\\
    Nielsen & 12,800 & 1,302 & 15.1M & 1,179 & 0.21\\
  \bottomrule
\end{tabular}
}
\end{table}

\textbf{Data Synthesis:}
Verifying a novel method for capturing item-item relationships in an empirical setting is challenging due to the lack of ground truth. Often these relationships are not known and could vary drastically from user to user. Owing to these issues,
most studies primarily evaluate the face validity of the model \cite{kang2018self, chen2018sequential}. The synthetic data allows us to test various effects and evaluate our model under a controlled environment. We start by examining our proposed model on synthetic data, comparing it to benchmarks, and evaluating the characteristics of the model (e.g., personalized and universal item-item relationships). Retail market basket data is one of the well-studied areas \cite{bordalo2013salience}. We follow common practice in synthesizing sessions to recreate heterogeneous user behaviors and various item-item relationships \cite{gabel2019p2v, manchanda1999shopping}. We extract personalized user behaviors from our model using only the basket data, withholding any information regarding items' relations or user choice models. We compare the extracted user behaviors to that of known values in our simulation. The simulation results validate the model hypothesis and provide an additional basis for performance under empirical data, where results are similar and consistently outperform state-of-the-art models. To the best of our knowledge, no proposed sequential recommender model provides a validated study on personalized item-item relationship. We next discuss the data generation schema.

The simulation of baskets is as follows: for a basket, $S^u_t$, with size $n^u_t$ purchased by user $u$ at time $t$, first the user chooses $n^u_t$ categories from a set of available categories. Then from each chosen category the user chooses a product $j$. A category $c$ is a set of similar items (e.g., various types of cereal, milk) and categories are disjoint. For simplicity, each basket can contain at most one item from a category and not all categories need to be purchased in a single basket. We use the multivariate normal distribution for simulating category choice model to capture various types of category-category relationships (e.g., complements vs. substitutes):
\begin{equation}\label{eq:cat}
    p^u_{ct} = \alpha_c + \epsilon^u_{ct}
\end{equation}
\noindent where $p^u_{ct}$ is probability of purchasing from category $c$ by user $u$ at time $t$, $\alpha_c$ is a constant category specific utility, and $\epsilon^u_{ct} \sim \mathcal{N}(0, \Sigma)$. The user chooses $n^u_t$ categories with highest probability to purchase from at time $t$. Next, we use multinomial probit model \cite{greene2003econometric} to simulate product choice within each category, explained in Appendix \ref{sec:syn_detail}. 
A positive value in the covariance matrix $\Sigma$ indicates two categories are purchased together frequently (e.g., milk and cereal), while a negative value represents contrary of that (e.g., fresh vs. frozen meat). We manually chose a block diagonal form for the covariance matrix $\Sigma$. Each block represents categories that have relations, while categories from separate blocks are independent of each other. We chose various sizes for different blocks (2-3-4) to represent low-mid-high order relationships between categories. For off-diagonal values of each block, we manually chose various positive/zero/negative values to represent complementary/coincidence/substitute product relationships. Further, we divide users into multiple disjoint groups. Each group follows a different $\Sigma$ to illustrate a deterministic user heterogeneity on top of the unknown user specific choice error terms. Heatmaps of $\Sigma$ are illustrated in Fig. \ref{fig:sim_results}. 
We synthesize a less dense dataset compared to the real-world data with 8,192 users, 2,000 items, and 5.8M million user-item interactions. Data statistics are summarized in Table \ref{tab:data_stats}. Further details of parameter values are presented in Appendix \ref{sec:syn_detail}.

\textbf{Empirical:} We use the Nielsen Consumer Panel\footnote{\url{https://www.chicagobooth.edu/research/kilts/datasets/nielseniq-nielsen}} dataset. The dataset comprises a representative panel of households. It includes all household purchases (from any outlet) intended for personal and at-home use \cite{nielsen}. This is a particularly challenging dataset in terms of sequential recommendation. Households are geographically dispersed over all states and demographically balanced to accurately represent the market in each area. The sessions are from different retailers. Over 4.3 million products in the dataset cover a variety of items including groceries, health and beauty aids, and alcohol. The majority of sequential recommenders evaluate their models on a limited set of users (e.g., within a state), a single retailer (e.g., a clothing retail or content provider), and a related set of items (e.g., all movies, or all clothing). We use data from 2014\textasciitilde2019 and randomly select 12,800 actively participating users with 15M+ user-item interactions from 50,000+ retailers. The products are grouped into hierarchical categories: 10 departments $\rightarrow$ 118 groups $\rightarrow$ 1,305 modules $\rightarrow$ 4.3+ million UPCs. We use \emph{Module} level as items in our experiment to reduce the sparsity of purchase patterns, and will refer to it as products in the rest of the paper. We keep one copy of the same item purchased in multiple quantities in each basket. We exclude items and users that have less than 10 records in the entire dataset. The statistics of data are summarized in Table \ref{tab:data_stats}.

\subsection{Experimental Setup}
The optimal architecture of PARSRec includes $N=1$ attention layer with $h=2$ heads. We observe that adding extra attention layers does not increase the performance significantly, presumably because the session lengths are short and a single transformer is able to induce all the necessary relationships within a session. Item and user embeddings are the size of $d_v=d_u=128$. We initialize the item and user  embedding matrices with uniform distributions of range $[-\frac{1}{\sqrt{|V|}}, \frac{1}{\sqrt{|V|}}]$ and $[-\frac{1}{\sqrt{|U|}}, \frac{1}{\sqrt{|U|}}]$, respectively. Other matrix parameters with size $n \times m$ are initialized with $\sim \mathcal{N}(0, \frac{2}{n + m})$. Dropout rate is set to 0.1 and mini-batch size is 256. We used Adam and Sparse Adam optimizers with learning rate 0.001, $\beta_1=0.9$, and $\beta_2=0.999$. The gradient is clipped at 30. We use PyTorch=v1.10.1 to implement the model\footnote{Codes for PARSRec: \url{https://github.com/ehgh/PARSRec}}.

\subsection{Evaluation Metrics}\label{subsec:metrics}
We use users' most recent session $S^u_{t_{|S_u|}}$ for testing, their second to last session $S^u_{t_{|S_u|}-1}$ for validation, and the rest for training. Note that the test set will have validation in its historical data as well. We predict items in the user's test basket sequentially (basket size > 1). For consistency, we adopt a similar strategy to \cite{kang2018self, sun2019bert4rec, he2017neural, tang2018personalized}, where they use a set of randomly sampled negative items plus ground truth item for evaluation purposes. We use a set of 100 randomly sampled items (negative and positive) plus ground truth items and rank these items to get the highest recommendations for the next item. Negative sampling is suitable in applications where users are less likely to interact with an item more than once (e.g., watching movies). However, in applications where users frequently interact with an item (e.g., grocery shopping, listening to music), evaluating the model on a pool of negative items (items that the user never interacts with) would make the prediction task more trivial. We observe that in the real-world dataset, about 81\% of test basket items also appear in the user's purchase history. We include positive items in the sampling to resolve this issue. 

For evaluation metrics, we report common top-N metrics \cite{kang2018self, sun2019bert4rec}: Hit Ratio (HR@k) and Normalized Discounted Cumulative Gain (NDCG@k) with $k \in  \{1, 5, 10\}$. NDCG is a rank-aware metric that penalizes the lower-ranked recommendations. Note that NDCG@1 equals HR@1. Additionally, we report another metric for session-level evaluation: session precision (\textbf{Sess-Prec@k}), the average ratio of items predicted accurately within a session:
\begin{equation}\label{eq:hitrate}
    session\ precision = \frac{|\{{\hat{v}_j | \hat{v}_j \in S^u_t}\}|}{|S^u_t|}
\end{equation}
\noindent In other words, session precision is the number of items that are recommended accurately normalized by the basket size.
\subsection{Benchmarks}
To validate the effectiveness of our proposed model, we benchmark it against the following baselines:
\begin{itemize}
    \item \textbf{POPRec}: A basic benchmark that recommends items based on their frequency of interactions by users.
    \item \textbf{SASRec} \cite{kang2018self}: It uses one-directional transformers for sequential recommendation. We modified negative sampling to sampling explained in Section \ref{subsec:metrics} for fair comparison.
    \item \textbf{BERT4Rec} \cite{sun2019bert4rec}: It applies bi-directional transformers with Cloze objective to sequential recommendation, and achieves state-of-the-art performance on many datasets.
\end{itemize}
We omit comparison with some other benchmarks like GRU4Rec \cite{hidasi2015session}, GRU4Rec$^+$ \cite{hidasi2018recurrent}, BPR \cite{rendle2012bpr}, NCF\cite{he2017neural}, and FPMC \cite{rendle2010factorizing} that have been outperformed by above methods on various real-world datasets \cite{kang2018self, sun2019bert4rec}. We use the code provided by the corresponding authors for benchmark models. We consider latent dimension sizes $d_v \in \{16, 32, 64, 128, 256\}$ for methods that include embeddings. The dropout rate is chosen from $\{0, 0.1, ..., 0.5\}$. We set $N=820$ (full sequence length) for synthetic dataset and $N=1000$ ($\sim$median of sequence length) for Nielsen dataset in SASRec and BERT4Rec that require setting sequence length. We either tuned all hyper-parameters on the validation set or referred to benchmarks' authors suggestions for optimal values. The reported results are under their optimal set of hyper-parameter values. 
\begin{table}
  \caption{Performance comparison of all benchmarks. Boldface and underlined values in each row represent the best and second best performances, respectively. Improvements of best to second best model are presented in the last column.}
  \label{tab:benchmarks}
  \resizebox{\columnwidth}{!}{%
  \begin{tabular}{llrrrrr}
    \toprule
    Dataset & Metric & POPRec & SASRec & BERT4Rec & PARSRec & Improvement\\
    \midrule
    \multirow{8}{*}{Synthetic} & HR@1 & 0.0005 & 0.1816 & \underline{0.1963} & \textbf{0.2136} & 8.8\%\\
      & HR@5 & 0.0028 & 0.3919 & \underline{0.4150} & \textbf{0.4791} &  15.4\%\\
      & HR@10 & 0.0066 & 0.5133 & \underline{0.5441} & \textbf{0.5825} &  7.0\%\\
      & NDCG@5 & 0.0013 & 0.2872 & \underline{0.3061} & \textbf{0.3489} &  13.9\%\\
      & NDCG@10 & 0.0026 & 0.3253 & \underline{0.3426} & \textbf{0.3800} &  10.9\%\\
      & Sess-Prec@1 & 0.0005 & 0.1826 & \underline{0.1974} & \textbf{0.2249} &  13.9\%\\
      & Sess-Prec@5 & 0.0029 & 0.3931 & \underline{0.4163} & \textbf{0.4784} &  14.9\%\\
      & Sess-Prec@10 & 0.0068 & 0.5151 & \underline{0.5357} & \textbf{0.5792} &  8.1\%\\
    \midrule
    \multirow{8}{*}{Nielsen} & HR@1 & 0.0008 & 0.1663 & \underline{0.1761} & \textbf{0.2444} & 38.8\%\\
      & HR@5 & 0.0921 & 0.4235 & \underline{0.4833} & \textbf{0.5934} &  22.7\%\\
      & HR@10 & 0.1620 & 0.5771 & \underline{0.6580} & \textbf{0.7355} &  11.7\%\\
      & NDCG@5 & 0.0454 & 0.2991 & \underline{0.3290} & \textbf{0.4208} &  27.9\%\\
      & NDCG@10 & 0.0680 & 0.3497 & \underline{0.3987} & \textbf{0.4632} &  16.1\%\\
      & Sess-Prec@1 & 0.0008 & 0.1669 & \underline{0.1856} & \textbf{0.2753} &  48.3\%\\
      & Sess-Prec@5 & 0.0919 & 0.4197 & \underline{0.4790} & \textbf{0.6093} &  27.2\%\\
      & Sess-Prec@10 & 0.1632 & 0.5729 & \underline{0.6344} & \textbf{0.7439} &  17.2\%\\
  \bottomrule
\end{tabular}
}
\end{table}
\begin{figure*}[ht!]
     \centering
     \begin{subfigure}[t]{.4\columnwidth}
         \centering
         \includegraphics[width=\textwidth]{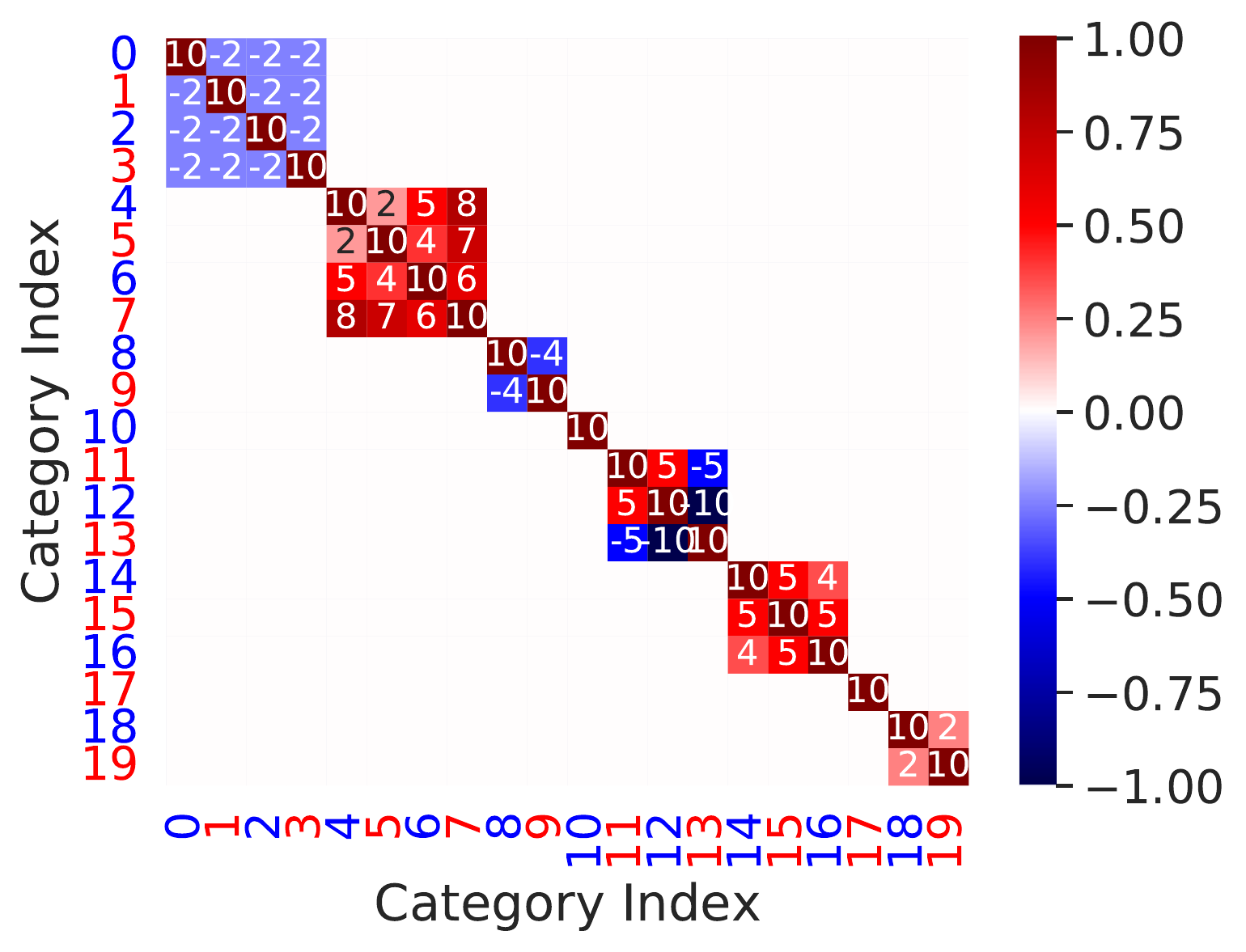}
         \caption{Average category purchase covariance matrix $\Sigma$ of all users}
         \label{fig:cov_full}
         \Description{Covariance matrix of category purchases}
     \end{subfigure}
     \hfill
     \begin{subfigure}[t]{.4\columnwidth}
         \centering
         \includegraphics[width=\textwidth]{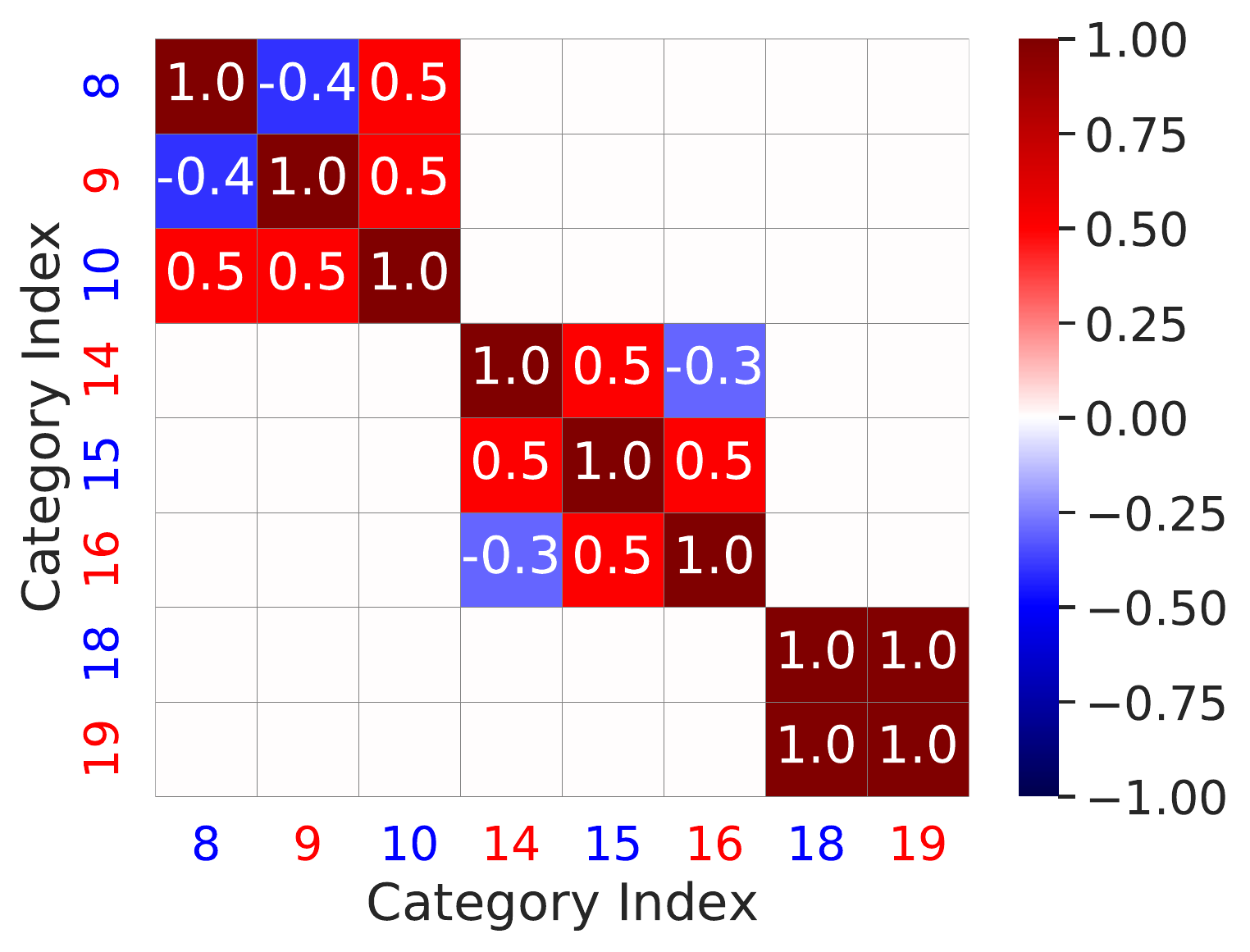}
         \caption{A submatrix of $\Sigma_A$ of user group $A$}
         \label{fig:cov1}
      \Description{A submatrix of covariance matrix for user group one}
     \end{subfigure}
     \hfill
     \begin{subfigure}[t]{.4\columnwidth}
         \centering
         \includegraphics[width=\textwidth]{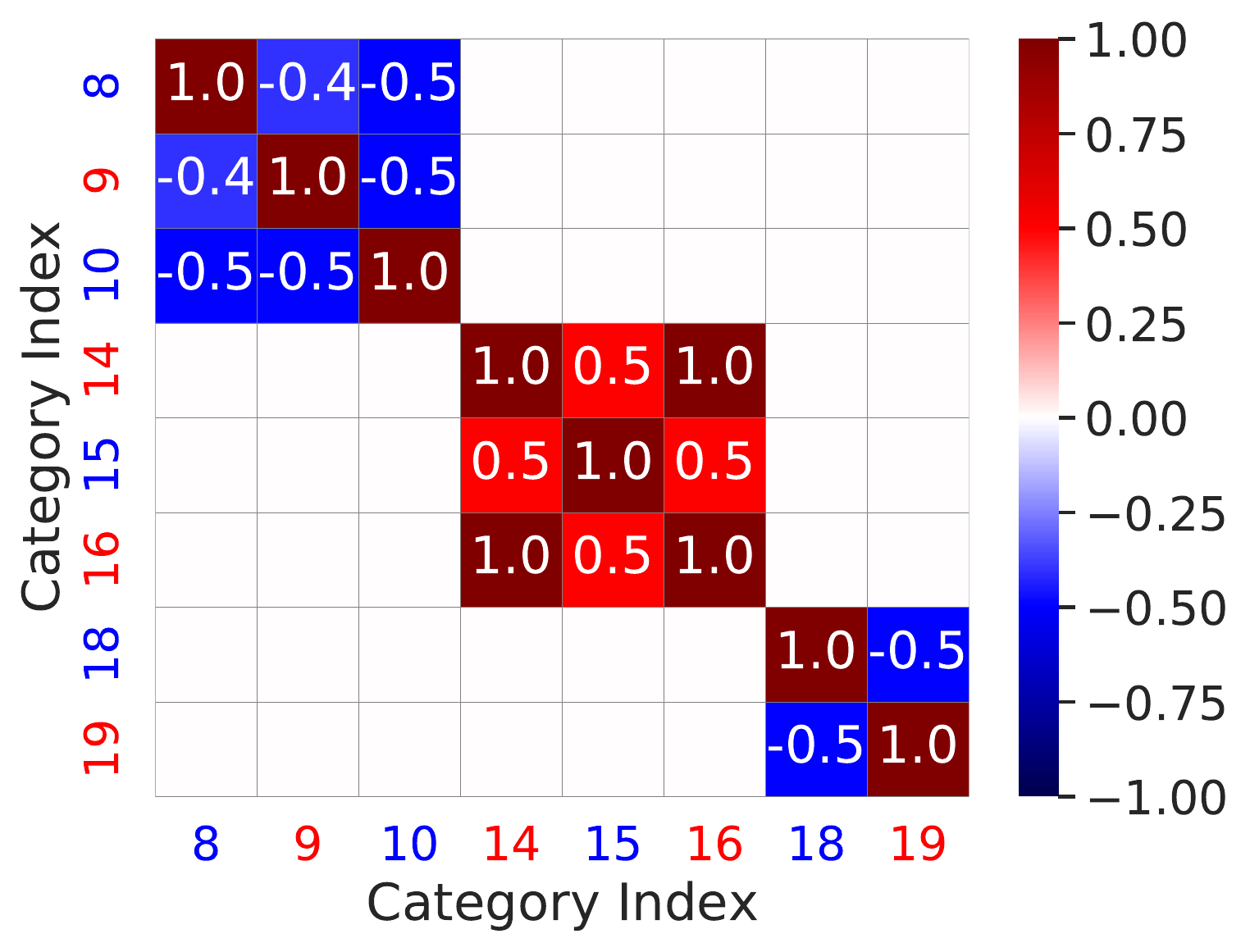}
         \caption{A submatrix of $\Sigma_B$ of user group $B$}
         \label{fig:cov2}
      \Description{A submatrix of covariance matrix for user group two}
     \end{subfigure}
     \hfill
     \begin{subfigure}[t]{.4\columnwidth}
         \centering
         \includegraphics[width=\textwidth]{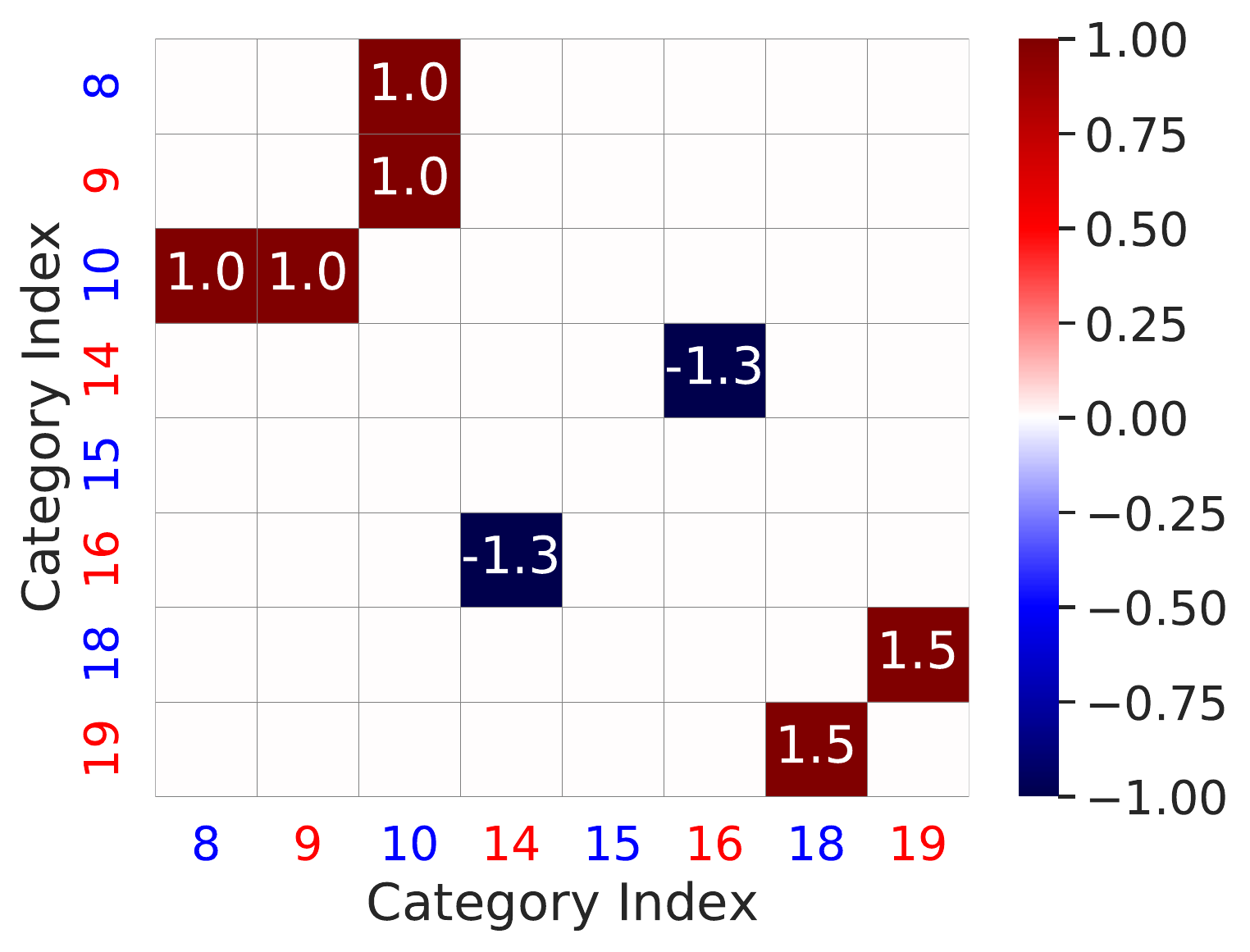}
         \caption{$\Sigma_A-\Sigma_B$, difference of user groups $A$ and $B$}
         \label{fig:cov_diff}
      \Description{Difference in the subset of covariance matrix between two user groups}
     \end{subfigure}
     \hfill
     \begin{subfigure}[t]{.4\columnwidth}
         \centering
         \includegraphics[width=\textwidth]{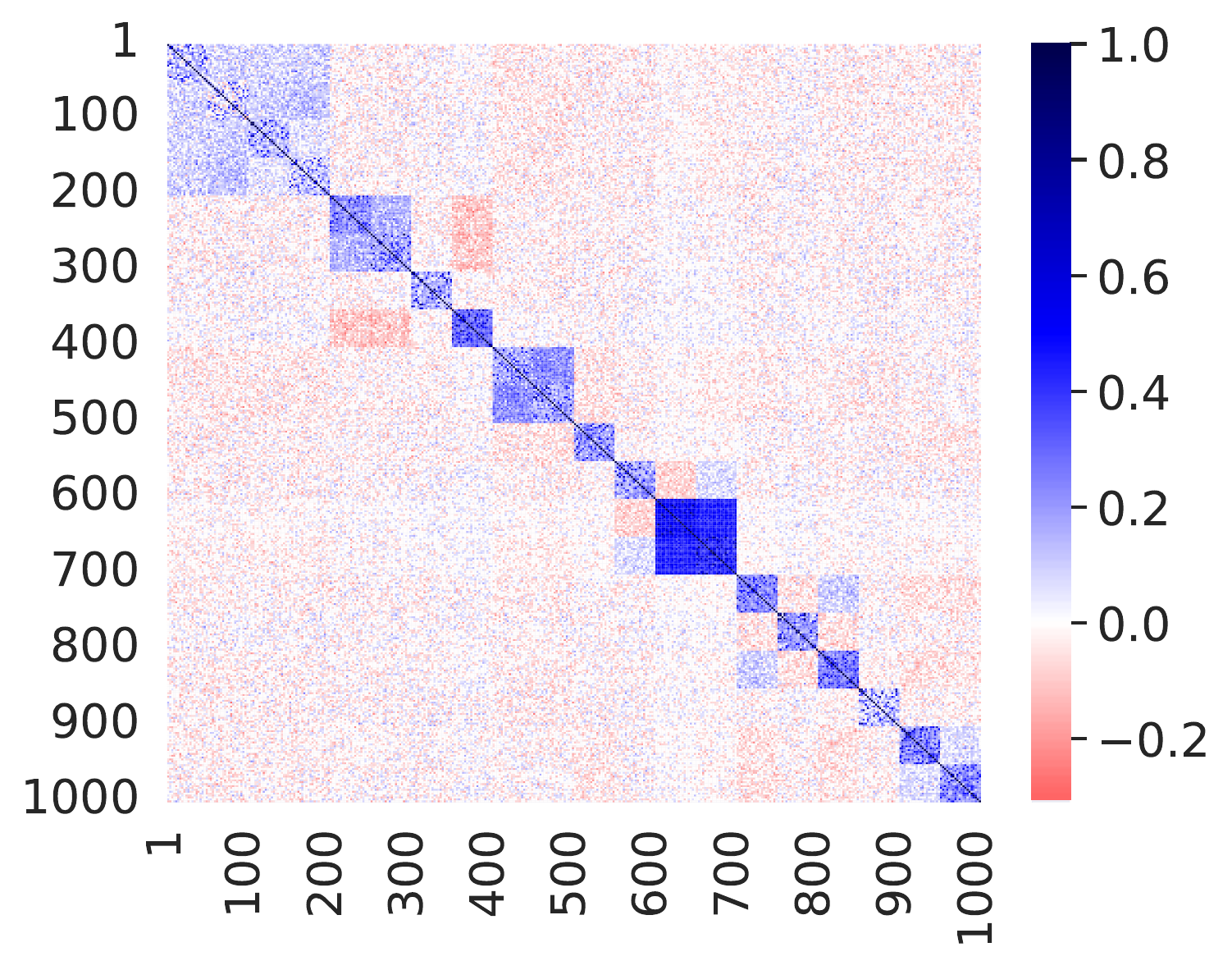}
         \caption{Item embeddings self-dot product heatmap ($\textbf{E}^V\cdot{\textbf{E}^{V^T}}$)}
         \label{fig:embedding}
      \Description{Category attention weights difference of two user groups}
     \end{subfigure}
     \hfill
     \begin{subfigure}[t]{.4\columnwidth}
         \centering
         \includegraphics[width=\textwidth]{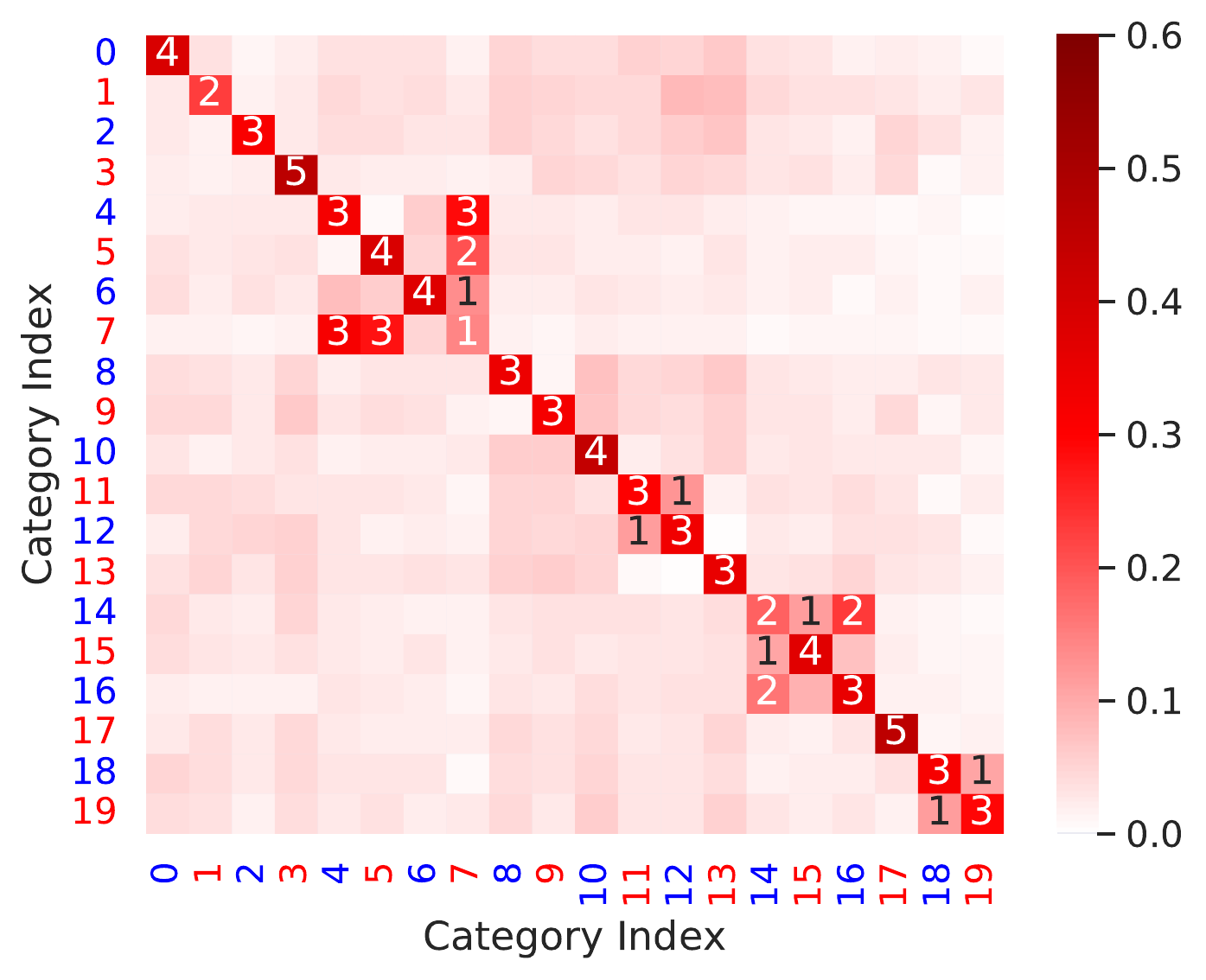}
         \caption{Average category attention weights learned by PARSRec}
         \label{fig:att_full}
         \Description{Average category attention weights learned by ParsRec}
     \end{subfigure}
     \hfill
     \begin{subfigure}[t]{.4\columnwidth}
         \centering
         \includegraphics[width=\textwidth]{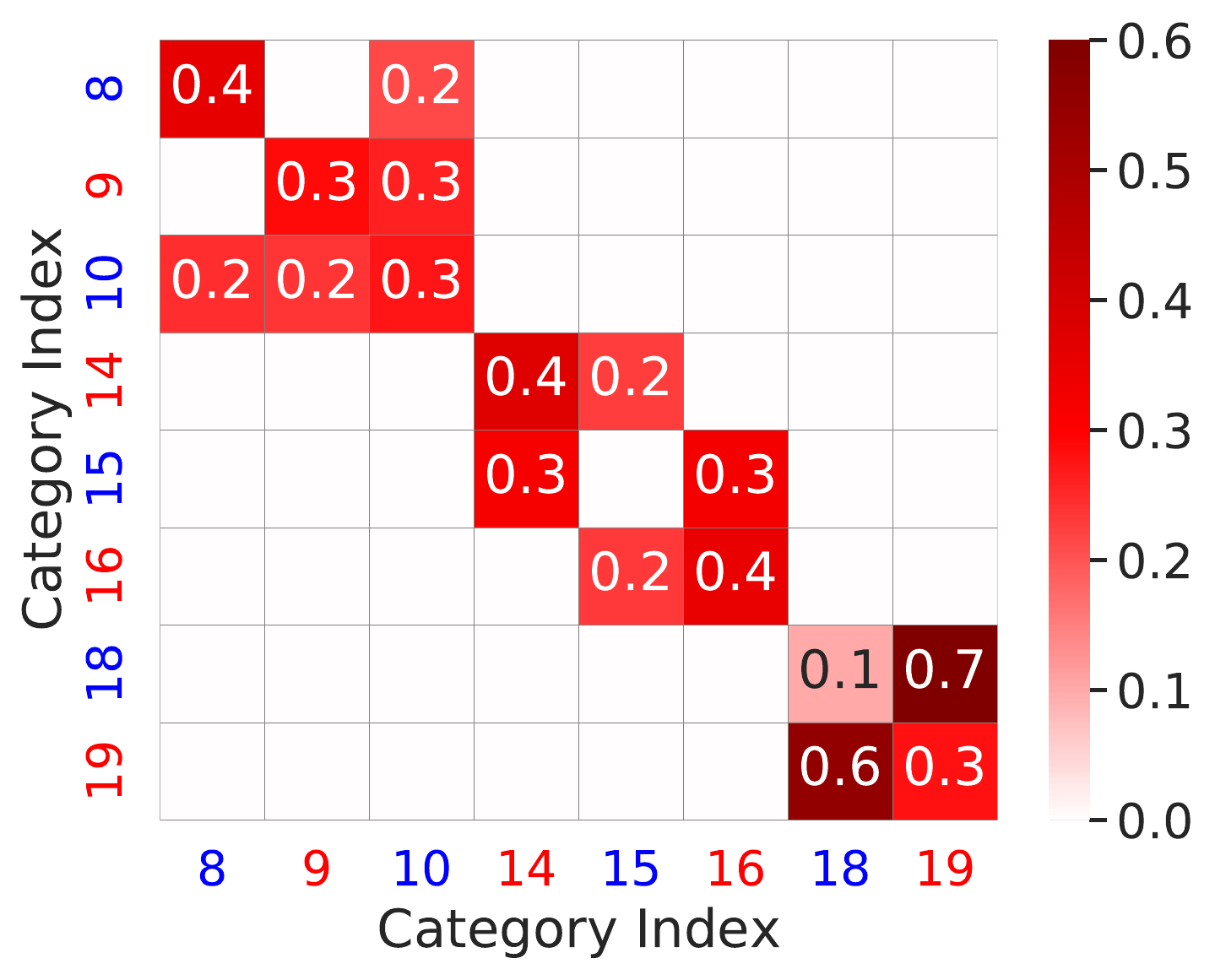}
         \caption{Attention weights of corresponding categories in (\ref{fig:cov1}) of user group $A$}
         \label{fig:att1}
      \Description{Attention weights of categories of user group one}
     \end{subfigure}
     \hfill
     \begin{subfigure}[t]{.4\columnwidth}
         \centering
         \includegraphics[width=\textwidth]{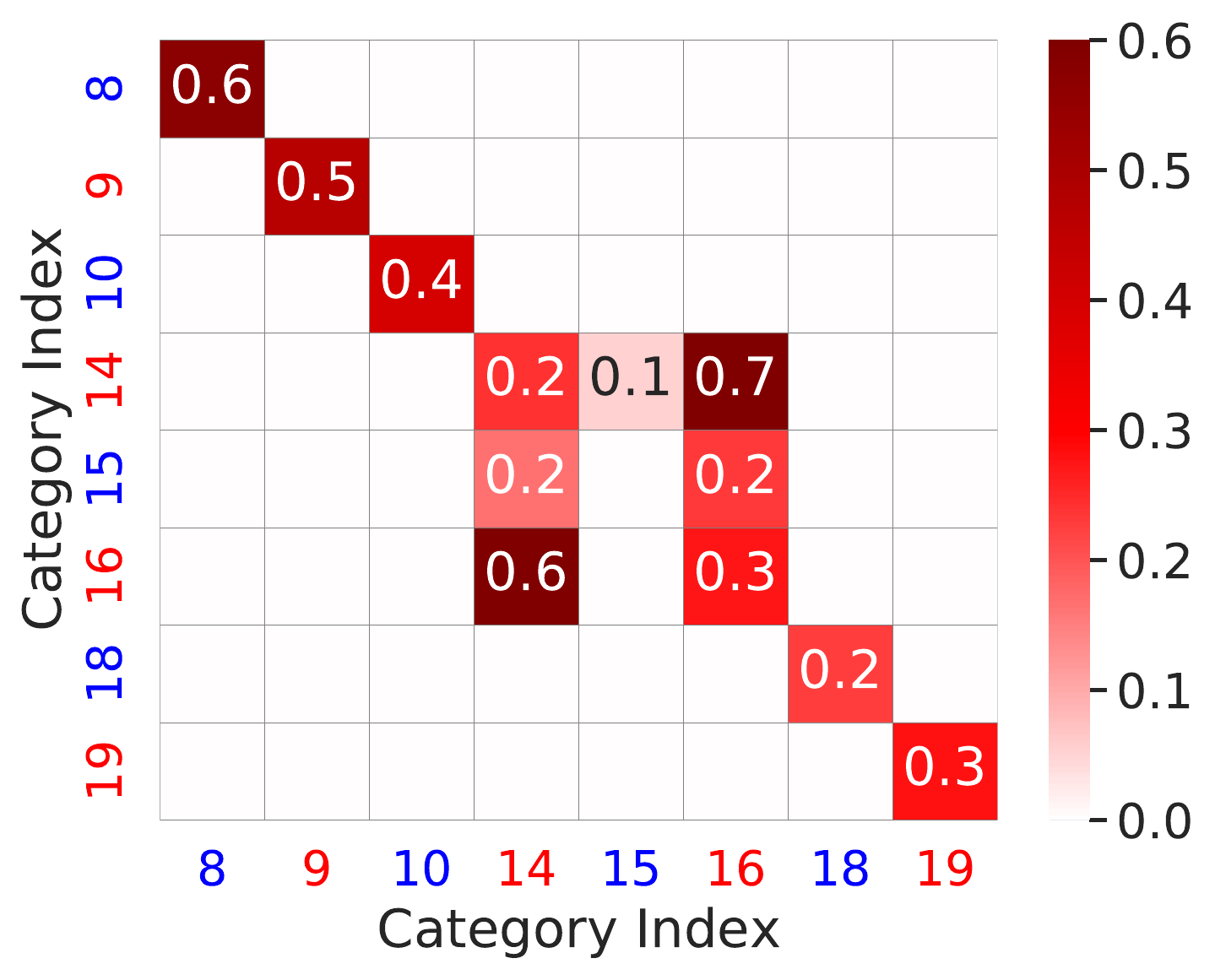}
         \caption{Attention weights of corresponding categories in (\ref{fig:cov2}) of user group $B$}
         \label{fig:att2}
      \Description{Attention weights of categories of user group two}
     \end{subfigure}     
     \hfill
     \begin{subfigure}[t]{.4\columnwidth}
         \centering
         \includegraphics[width=\textwidth]{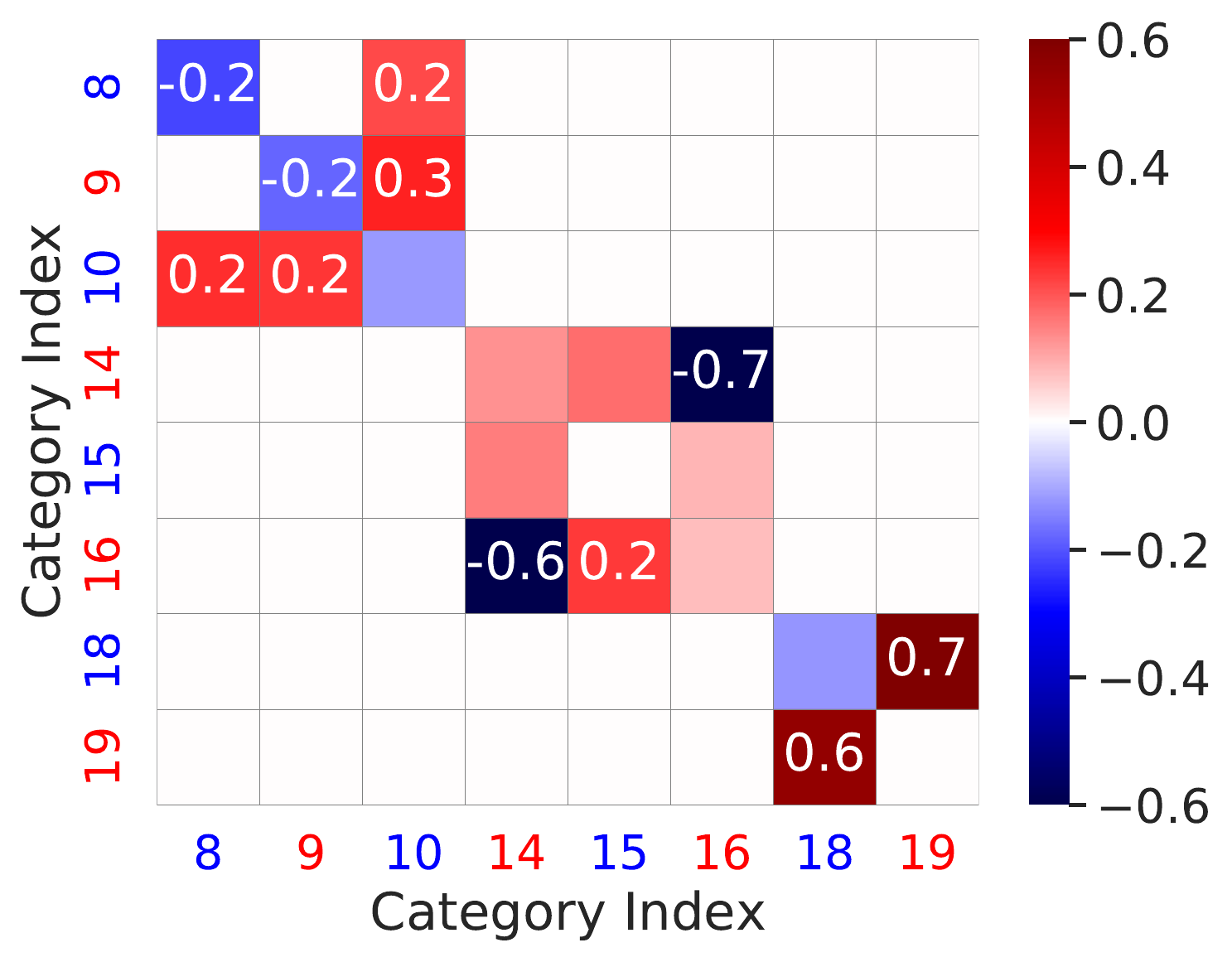}
         \caption{Category attention weights difference of user groups $A$ and $B$}
         \label{fig:att_diff}
      \Description{Category attention weights difference of two user groups}
     \end{subfigure}
     \hfill
     \begin{subfigure}[t]{.4\columnwidth}
         \centering
         \includegraphics[width=\textwidth]{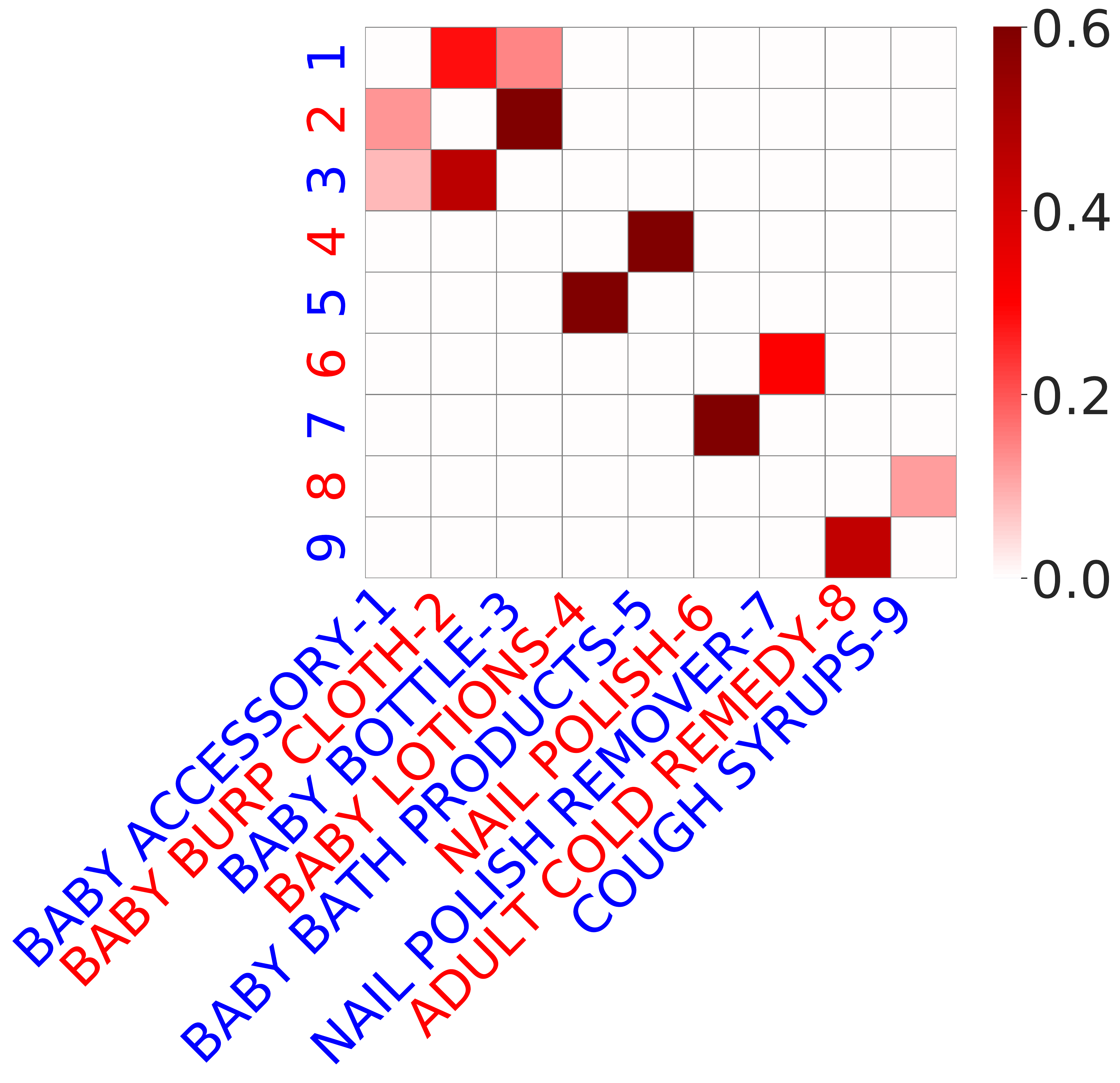}
         \caption{Empirical data: Average attention weights of items for users with babies}
         \label{fig:att_emp}
      \Description{Attention weights of sample items in empirical data}
     \end{subfigure}     
     \caption{Simulation and empirical heatmaps: (a) Average covariance matrix $\Sigma$ of category purchases $p^u_{ct}\sim \mathcal{N}(\alpha_c, \Sigma)$ in data synthesis. A positive value in $\Sigma$ indicates
two categories are purchased together frequently (e.g. milk and
cereal), while a negative value represents contrary of that. (b, c) A subset of covariance matrix for user groups $A$ and $B$, (d) difference of (b) and (c), (e) Item embedding self-dot-product shows embedding has learned global item categories and relationships. (f) Average category attention weights learned by the model. Higher values correspond to positive values in (a) while lower values correspond to zero/negative values in (a), (g, h) Category attention weights for selected categories in (b, c), the matching of (b<->g) and (c<->h) illustrates that model can learn personalized attention. Values < 0.05 are filtered for clarity, (i) difference between (g) and (h). The high values matches Fig. (d). (j) Attention weights of sample products in empirical data for a group of users with babies. The network learns correct relations without prior knowledge of users or products (Color represents sign and intensity represents value. Annotations in (a) and (f) are scaled by 10 for readability. All other annotations illustrate the true values of the cell. Figure best viewed in color. Axis label colors are for readability)}
     \label{fig:sim_results}
\end{figure*}

\textbf{Performance Comparison:} Table \ref{tab:benchmarks} summarizes the recommendation performance of all benchmarks as well as our proposed model on synthetic and Nielsen datasets. The last column shows the statistically significant improvements of the best model to the second-best model. We observe that POPRec has the lowest performance, presumably because it does not look into user-specific behaviors. SASRec uses transformers and performs lower than BERT4Rec because it only looks at item relationships from left to right. BERT4Rec also uses transformers but in a bidirectional setting that allows learning from both right and left. PARSRec performs consistently better than all methods on both datasets. It is likely because PARSRec differentiates between interactions within a session and interactions in the past sessions by using RNN blocks. It also uses explicit user queries in its attention blocks to learn personalized item-item relationships which is explained more in detail in Section \ref{subsec:discussion}. PARSRec gains an average of 19.1\% HR@5, 20.9\% NDCG@5, and 21.1\%  Sess-Prec@5 improvement over the second-best benchmark.

\subsection{Discussion}\label{subsec:discussion}

The key deliverable of our study was to develop a personalized recommendation system that accounted for patterns of individual preferences over time and item relationships. The model developed here can be applied to multiple scenarios, from populating song lists to basket completion exercises. The model's success hinges on its ability to address two key questions: (1) how does PARSRec capture personalized item-item relationships? and (2) how does learning personalized item-item relationships affect the recommendation?

\subsubsection{Extracting Personalized Item Relationships}
Recommender systems follow one of the two common methods to capture heterogeneous user behaviors: \emph{i)} learn an explicit user representation based on the user behavior \cite{tang2018personalized, rendle2010factorizing}; \emph{ii)} implicitly represent users by aggregating the embeddings of user interacted items \cite{kang2018self, sun2019bert4rec, hidasi2015session, hidasi2018recurrent, he2016fusing}. PARSRec takes the former approach by learning an explicit user embedding. Recently, the attention layers of transformers in the state-of-the-art recommender systems provide an explainable visual for item-item relationships. These models (e.g., BERT4Rec \cite{sun2019bert4rec}, and SASRec \cite{kang2018self}) often use a self-attention layer where all key, value, and query parameters consist of item embeddings (implicit user representation). Hence, they learn global item-item relationships that differentiate users via their purchase history. However, in PARSRec the query consists of explicit user embeddings (plus user purchase history, if available). This allows for the network to explicitly learn and visualize the personalized item-item relationships. To validate this capability, we conduct a controlled simulation on the synthetic dataset. We split users into various groups and each group has a different category covariance matrix, $\Sigma$, during data synthesis. \Cref{fig:cov_full,fig:cov1,fig:cov2,fig:cov_diff} illustrate the heatmaps (of a subset) of $\Sigma$ for two different user groups $A$, and $B$ (along with their average and difference). We then extract the highest attention weights of each user at every step of the training phase (averaged on attention heads). Note that we only extract weights of the valid non-padding items. We aggregate the item-item attention weights to category level for ease of comparison and readability. We perform row-wise sum followed by column-wise average within each category, then row-wise normalization over all items for each user individually. \Cref{fig:att_full,fig:att1,fig:att2,fig:att_diff} show the heatmaps of the category level attention weights for two different user groups $A$ and $B$, and their average and difference. Further, Figure \ref{fig:embedding} represents the heatmap of dot-product of item embedding with itself ($\textbf{E}^V\cdot{\textbf{E}^{V^T}}$). We observe following traits from the figures:
\begin{itemize}[noitemsep,topsep=0pt,leftmargin=*]
    \item (\ref{fig:cov_full}) vs. (\ref{fig:att_full}): We expect the attention weights in (\ref{fig:att_full}) to capture the average category relations for all users corresponding to Figure (\ref{fig:cov_full}). We observe that there is a correlation between attention weights and covariance matrices. Higher attention weights correspond to larger positive covariance values, and near-zero attention weights correspond to the negative covariance values. We also observe that attention weights capture category independence in the form of block diagonal matrices. Note that the network has no prior knowledge of item categories, and user-item interactions are the only information introduced to the network. 
    \item (\ref{fig:cov1}, \ref{fig:cov2}, \ref{fig:cov_diff}) vs. (\ref{fig:att1}, \ref{fig:att2}, \ref{fig:att_diff}): We observe that PARSRec can learn user heterogeneity by accurately extracting different user groups' attention weights. We see a one-to-one match between user groups' covariance matrices in (\ref{fig:cov1}, \ref{fig:cov2}) and their corresponding attention weights in (\ref{fig:att1}, \ref{fig:att2}). For easier comparison, we include their differences in (\ref{fig:att_diff}). The cells with higher absolute values in Figure (\ref{fig:att_diff}) match the differences in Figure (\ref{fig:cov_diff}). Note that the network has no prior knowledge of user groups and infers their difference via user behavior.
    \item (\ref{fig:cov_full}) vs. (\ref{fig:embedding}): The global item-item similarity (i.e. dot-product) matches the average relationships of items. We observe the similarity of items within each category illustrated by blue colors. This is because in data synthesis, a user interacts with at most one item per category during a session, and items within a category are expected to have similar embedding. (\ref{fig:embedding}) also shows a matching pattern with global category correlations where red cells match high positive category covariance values and blue cells match low positive or negative covariace values. Note that unlike attention weights, item embeddings are learned relative to other item embeddings. E.g., categories 14 and 16 show similarities in Figure (\ref{fig:embedding}) (blue color) even though they have a positive but relatively lower correlation compared to category pairs 14-15 and 15-16 (red color). However, attention weights of categories 14 and 16 are captured positively in Figure (\ref{fig:att_full}).  (\ref{fig:embedding}) shows that the model learns the item similarities globally to some extent. However, item representations alone cannot present personalized item relationships like the attention layer does.
\end{itemize}
\begin{figure}
     \centering
     \begin{subfigure}[t]{.49\columnwidth}
         \centering
         \includegraphics[width=\linewidth]{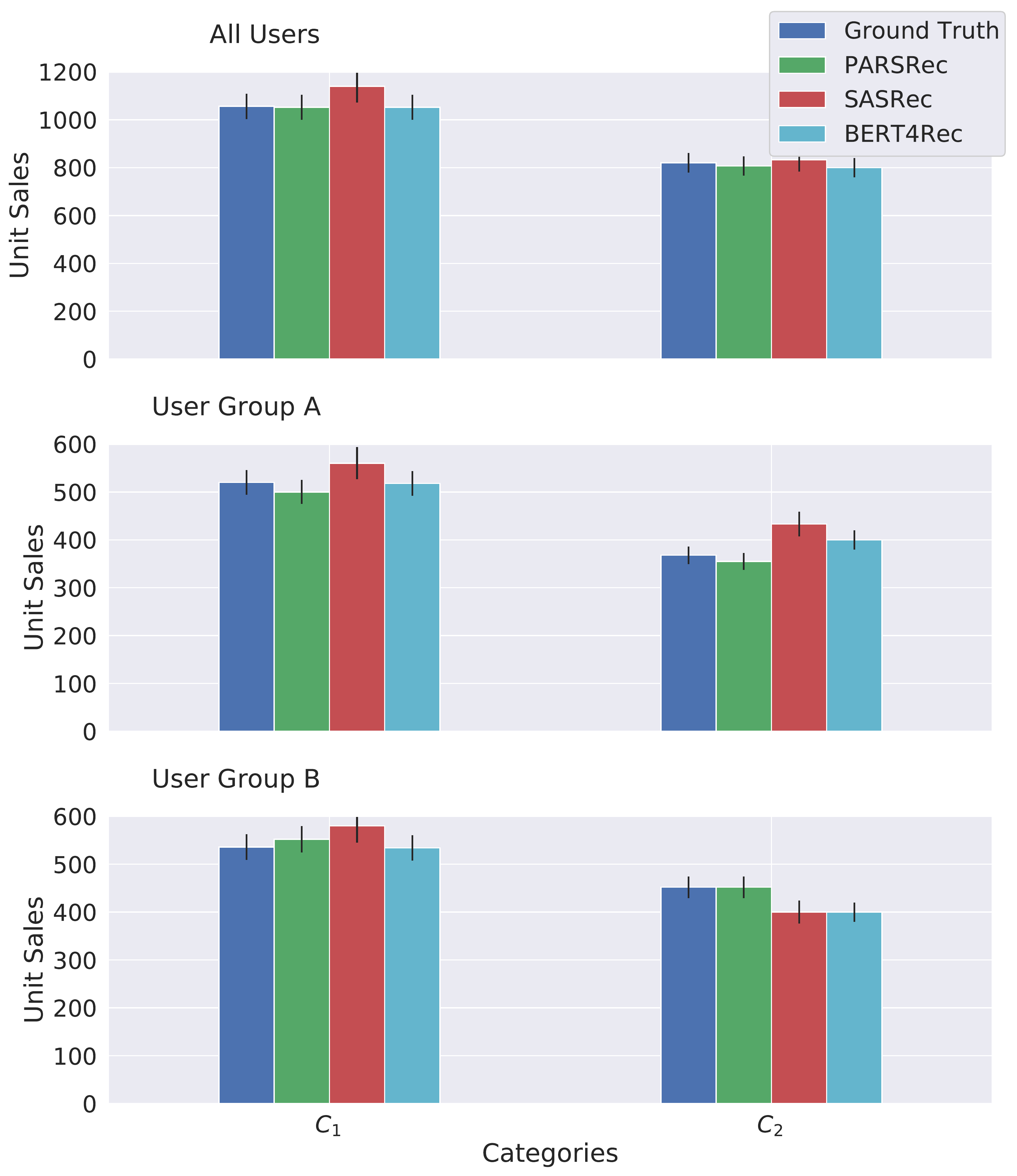}
         \caption{Sales estimation of different categories for user groups A and B}
         \label{fig:prediction_sales}
         \Description{estimating sales}
     \end{subfigure}
     \hfill
     \begin{subfigure}[t]{.49\columnwidth}
         \centering
         \includegraphics[width=\linewidth]{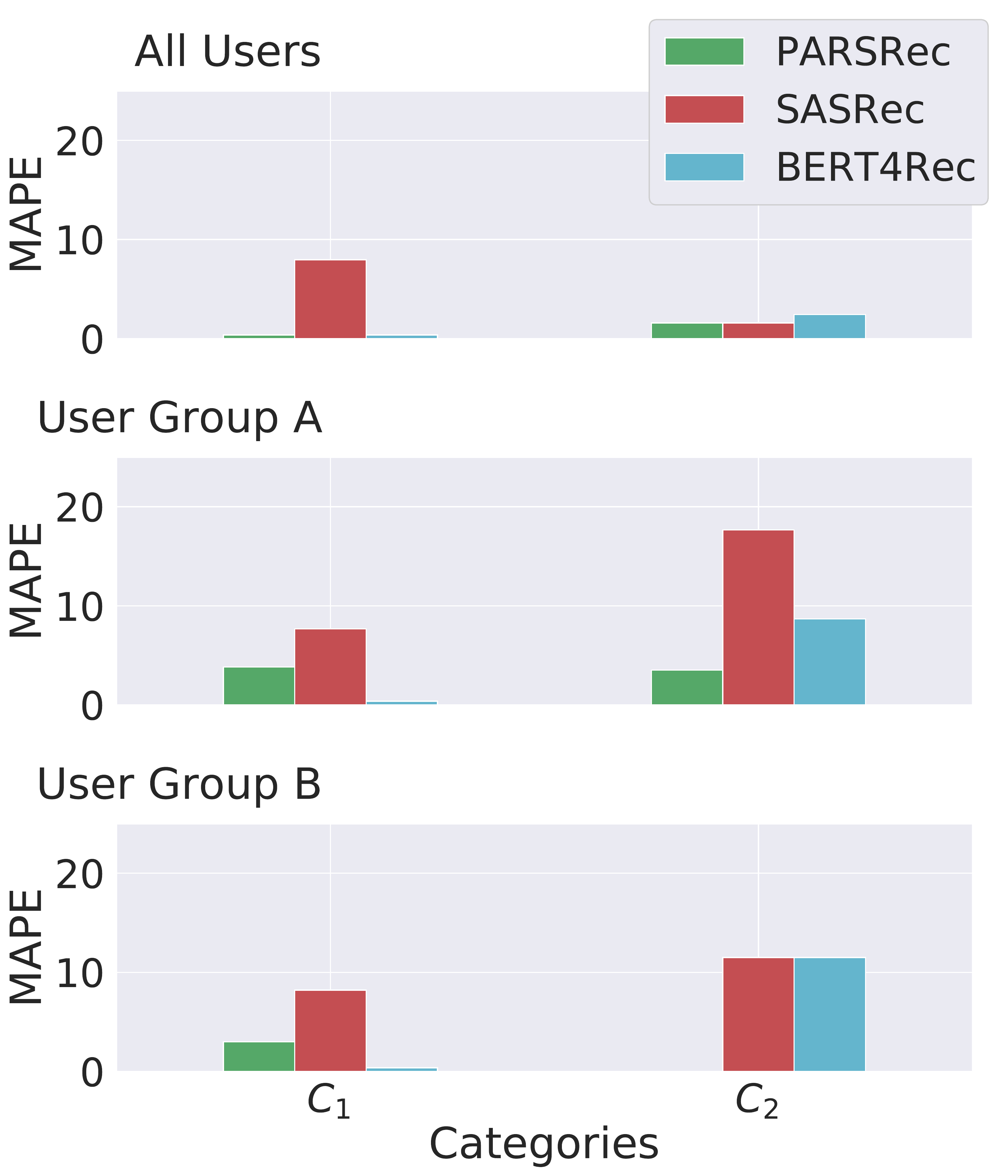}
         \caption{Sales estimation MAPE}
         \label{fig:mape}
      \Description{Sales estimation MAPE}
     \end{subfigure}
     \caption{Estimating the influence of removing category $C_0$ (present at training phase) from the assortment at testing phase on sales of categories $C_1$ and $C_2$ for two user groups. Cov($C_0$,$C_2$)=0.6 for user group A. Other covariances are zero. PARSRec captures category spillover effects.}
     \label{fig:sales}
\end{figure}
\textbf{Empirical Attention Between Items:} The empirical dataset lacks the ground truth for user behavior. However, we can look into item relationships learned by the network. Figure (\ref{fig:att_emp}) shows the average of a subset of attention weights between items for a group of users who have babies in the Nielsen dataset. We observe that the network learns to separate different product groups related to each other without prior knowledge of users or item categories. The network also identifies subgroups of related items within a category (e.g., \{baby bath products, baby lotion\} vs. \{baby bottle, baby burp cloth, baby accessory\}) even though they occur in the same sessions.
\subsubsection{Effect of Personalized Recommendation}
To investigate the importance of learning personalized item-item relationships in making recommendations, we conduct a controlled simulation on the synthetic dataset. We drop a product category from the pool of products and observe its consequences on other categories' sales. For example, it is observed when a retailer discontinues tobacco and cigarettes, they face a cross-category spillover effect on other products such as alcoholic beverages \cite{goli2021happens}. We mimic the real-world behavior in our simulation as follows: 1) we assume the assortment contains all products during training and validation phase, 2) we remove baskets that contain the dropped category $C_0$ from the \emph{test set} and also from the pool of products \emph{during testing phase}, and 3) we estimate the unit sales of two types of categories for two user groups during testing phase using top 10 recommendations. Category $C_1$ is independent of $C_0$ for both user groups. Category $C_2$ is highly correlated with $C_0$ for user group $A$ (cov=0.6), but independent of $C_0$ for user group $B$ (cov=0). Figure \ref{fig:prediction_sales} shows the estimates of category sales for each user group, and Figure \ref{fig:mape} shows Mean Absolute Percentage Error (MAPE) of sales. We observe that PARSRec accurately distinguishes the spill over effect of $C_0$ removal on categories $C_1$ and $C_2$. PARSRec estimates the cross-category drop in sales of category $C_2$ for user group $A$ (MAPE=3.5\%) and uninfluenced sales of $C_2$ for user group $B$ (MAPE=1\%). It also accurately estimates the sales of the independent category $C_1$ for both user groups (MAPE=3.8\% and 3.7\%). It is important to note that all training/validation sets include category $C_0$ and only the test set is subject to change. Without any prior training on category removal effects, the proposed model can predict the user behavior by learning personalized item-item relationships. This allows retailers to conduct simulated experiments on assortment modifications and estimate their effect on different customers without the costs of applied experiments.

\section{Ablation Study}\label{sec:app-ablation}
\begin{table}
  \caption{Ablation analysis (NDCG@10 and Sess-Prec@10) of synthetic and empirical dataset. An increase in performance from the default PARSRec setting is emphasized in bold.}
  \label{tab:ablation}
  \resizebox{\columnwidth}{!}{%
  \begin{tabular}{lrrrr}
    \toprule
    \multirow{2}{*}{Architecture} & \multicolumn{2}{c}{Synthetic} & \multicolumn{2}{c}{Nielsen} \\  \cline{2-3}\cline{4-5}
     & Sess-Prec@10 & NDCG@10 & Sess-Prec@10 & NDCG@10\\
    \midrule
    PARSRec default & 0.5792 & 0.3800 & 0.7439 &  0.4632 \\
    Remove LN & 0.5756 & 0.3746 & 0.6775 & 0.4183  \\
    Remove Dropout  & 0.5780 & 0.3775 & 0.7356 & 0.4553 \\
    Remove Q in LN & 0.5643 & 0.3745 & 0.7400 & 0.4601 \\
    \#Att Layers=2 & 0.5506  & 0.3622 & 0.7320 & 0.4543  \\
    \#heads=1  & 0.5769 & 0.3781 & 0.7399 & 0.4602 \\
    \#heads=4  & 0.5790 & 0.3799 & 0.7431 &  0.4628 \\
    FFN pre-RNN  & 0.5723 & 0.3775 & 0.7426 &  \textbf{0.4637} \\
    FFN post-RNN  & 0.5009 & 0.3362 & 0.6900 & 0.4189 \\
    Extra Features  & - & - & \textbf{0.7591} & \textbf{0.4759} \\
    
  \bottomrule
\end{tabular}
}
\end{table}
We conduct an ablation study to understand the impact of various components of PARSRec and some variations in the design of the network. 
Table \ref{tab:ablation} summarizes the performance of the optimal model and eight variants on both synthetic and empirical datasets. All other hyper-parameters are unchanged. The variants are as follows:

(1,2) \emph{Remove LN, Dropout}: Including both components help improve the performance, with LN being more prominent on empirical dataset. 

(3) \emph{Remove \textbf{Q} from LN}: Adding \textbf{Q} at LN increases the accuracy. It carries the user history and is beneficial to add it to RNN input. 

(4)  \emph{Number of Attention Layers}: Stacking more attention layers achieves similar performance on the empirical dataset but performs lower on the synthetic dataset. Presumably, because the sessions are short and single layer attention is sufficient to learn the relationship of items within a session. Longer sequences might pose more complex patterns and require more layers.

(5) \emph{Number of heads in Multi-Head Attention}: The results show that increasing the number of heads beyond two does not significantly boost performance, presumably because of small session lengths. Longer sessions might benefit from more number of heads.

(6) \emph{FFN pre-RNN input}: We explore adding layers of Feed-Forward Network right before the input of RNN. The results show that the added FFN has a similar performance. It is likely that the projection layer of the attention mechanism extracts the required features, and adding extra layers is redundant.

(7) \emph{FFN post-RNN output}: Adding FFN after RNN output significantly decreases the performance, possibly due to overfitting. The optimal dimension of RNN is satisfactory for output prediction.

(8) \emph{Extra User Features}: We explore adding extra user and session features from the empirical dataset. We pass continuous features through a Multi-Layer Perceptron and further concatenate them with the user and categorical features' embeddings to use as the initial hidden state. Features include income, age, gender, location, retailer id, day of the week of session, household size and composition, user education, marital status, and race. These features contribute slightly to improving performance ($\sim$2\% increase). This confirms that the personalized model learns user features that contribute to the recommendation task via user embedding.
\section{CONCLUSION AND FUTURE WORK}
We proposed PARSRec, a sequential recommender model that combines attention mechanism and RNN. PARSRec learns personalized item relationships by using explicit user embeddings in the query of attention mechanism. We conduct a controlled simulation on a synthetic dataset to validate user behavior learning. Empirical results on Nielsen's Consumer Panel dataset show that PARSRec outperforms state-of-the-art self-attention models. One future direction is incorporating the attention mechanism with GRU/LSTM networks for longer sessions and exploring the impact of a different loss function such as Bayesian personalized ranking. Another direction would be to leverage item features (e.g., textual information, price, flavor) in the attention mechanism.

\begin{acks}
“Researcher(s)' own analyses calculated (or derived) based in part on data from Nielsen Consumer LLC and marketing databases provided through the NielsenIQ Datasets at the Kilts Center for Marketing Data Center at The University of Chicago Booth School of Business.”
“The conclusions drawn from the NielsenIQ data are those of the researcher(s) and do not reflect the views of NielsenIQ. NielsenIQ is not responsible for, had no role in, and was not involved in analyzing and preparing the results reported herein.” The authors thank Erin Nannery, Ashutosh Nayak, Jörn Boehnke, and Kourosh Vali for their help and assistance.
\end{acks}

\bibliographystyle{ACM-Reference-Format}
\bibliography{references}

\appendix
\section{Data Synthesis}\label{sec:syn_detail}
We use multinomial probit model \cite{greene2003econometric} to simulate product choice within each category, $j^{u,c}_t$:
\begin{equation}\label{eq:utility}
\begin{aligned}
    j^{u,c}_t &= \argmax_{j \in c}{\eta^{u,c}_{jt}}\\
    \eta^{u,c}_{jt} &= \omega^{u,c}_{j} - \beta^c \nu^c_j + \gamma^{u,c}_{jt}
\end{aligned}
\end{equation}
\noindent where $\eta^{u,c}_{jt}$ is utility of item $j$ in category $c$ for user $u$ at time $t$, $\omega^{u,c}_{j} \sim \mathcal{N}(0, \Omega^c)$ is the product base utility, $\beta^c \nu^c_j$ is a disutility for paying price $\nu^c_j$, and $\gamma^{u,c}_{jt} \sim \mathcal{N}(0, \sigma^c)$ is a random term to capture any uncontrollable parameter affecting the product choice. 
The simulation first chooses categories based on Eq. (\ref{eq:cat}) and the user chooses the product with highest utility in each category based on Eq. (\ref{eq:utility}). 
We use $C=20$ categories, each with $|V^c| = 100$ products. Basket size $n^u_t$ is sampled from Weibull(0.80, 1.47) to represent the basket size distribution in our real-world data. Single item baskets and large (>10) baskets (tail of the distribution) are filtered.
We set $\sigma^c=1, \beta^c=0.1$, and $\alpha_c=-0.5$. Category specific covariance matrix $\Omega^c=\tau_c^2 \Omega^c_0$ where $\tau_c=2$ is the standard deviation and $\Omega^c_0$ is the correlation matrix. $\Omega^c_0$ is a positive-semidefinite matrix generated using vine method \cite{lewandowski2009generating} under Beta(0.2, 1) distribution to simulate various degrees of product competition. Product prices $\nu^c_j \sim Uniform(\frac{\nu^c}{2}, 2\nu^c)$ where category base price $\nu^c \sim LogNormal(0.5, 0.1)$ is set to mimic the real-world data prices. 
\section{Mathematical Model of PARSRec Architecture}\label{sec:app-model}
\textbf{Embedding Layer}

\begin{itemize}
    \item We convert item indices of the basket $S^u_{t_i}=[v^{(S)}_1, ..., v^{(S)}_{|n^u_i|}]$ to input embedding vectors $\textbf{K}^u_{i} \in \mathbb{R} ^ {n^u_i \times d_v}$ where the $j$-th row of $\textbf{K}^u_{i}$ is $\textbf{E}^V_{v^{(S)}_j}$. We use subsets of $\textbf{K}^u_{i}$ as keys and values to attention layer. The details are explained in section \ref{ssec:attention}.
    
    \item We convert item indices of the previous sessions, $(S^u_{t_1}, ..., S^u_{t_{i-1}})$ to their embedding representations by $\textbf{E}^V$ and reduce them using the weighted sum operator to a single representation vector, $\textbf{H}^u_{i-1} \in \mathbb{R} ^ {1 \times d_v}$:
    \begin{equation}
        \textbf{H}^u_{i-1} = \sum \limits_{v \in \{w\in s | s \in \{S^u_{t_1}, ..., S^u_{t_{i-1}}\}\}}{\textbf{E}^V_v}
    \end{equation}
    \item We can merge user features $\textbf{F}^U_u$ with the user embedding $\textbf{E}^U_u$ to create a combined user input to our model:
    \begin{equation}
        \hat{\textbf{E}}^U_u = combine(\textbf{E}^U_u, \textbf{F}^U_u)
    \end{equation}
    where $\hat{\textbf{E}}^U_u, \textbf{E}^U_u, \textbf{F}^U_u$ are combined input, embedding and features of user $u$, respectively, and $combine(.)$ is any function that merges two vectors into one. Examples are (weighted) addition, concatenation, and element-wise multiplication. In our ablation study, we investigated the choice of concatenation.
    
\end{itemize}

\vspace{2cm}
\textbf{Attention Layer}

The multi-head attention with $h$ heads can be summarized as:
\begin{align*}
    MultiHead&(\B{Q}, \B{K}, \B{V}) = Concat(head_1, ..., head_h)\B{W}^O\\
    head_i &= Attention(\B{Q}\B{W}^Q_i, \B{K}\B{W}^K_i, \B{V}\B{W}^V_i)\\
    &= softmax\Big(\frac{(\B{Q}\B{W}^Q_i)(\B{K}\B{W}^K_i)^T}{\sqrt{d_k}}\Big)(\B{V}\B{W}^V_i)
\end{align*}
\noindent where the projection matrices $\B{W}^O \in \mathbb{R}^{hd_v \times d_{v}}$, $\B{W}^Q_i \in \mathbb{R}^{d_{q} \times d_{q}}$, $\B{W}^K_i \in \mathbb{R}^{d_{k} \times d_q}$, and $\B{W}^V_i \in \mathbb{R}^{d_{v} \times d_q}$ are learned parameters, query $\B{Q}$, key $\B{K}$, and value $\B{V}$ are matrices, $d_k$ is the dimension of the key and value, and $d_q$ is the dimension of the query. Key, value, and query are:
\begin{gather*}
    \B{Q} = Concat(\textbf{E}^U_u, h_j)\\
    m\text{-th row}(\B{K}) = m\text{-th row}(\B{V}) = \textbf{E}^V_{v^{(S)}_m}
\end{gather*}
\textbf{Recurrent Layer}

The RNN and prediction layers combined are:
\begin{gather*}
    h_{j+1} = ReLU(\tilde{v}_j \textbf{W}^{(1)} + \tilde{h}_j \textbf{W}^{(2)} + \textbf{b}^{(1)}) \\
    \hat{v}_{j+1} = \argmax{((\tilde{v}_j \textbf{W}^3 + \tilde{h}_j \textbf{W}^4 + \textbf{b}^2).\textbf{E}^{V^T})}
\end{gather*}
\noindent where $\textbf{W}^1 \in \mathbb{R}^{d_v \times d_v}, \textbf{W}^2 \in \mathbb{R}^{(d_u+d_v) \times d_v},  \textbf{W}^3 \in \mathbb{R}^{d_v \times |V|}, \textbf{W}^4  \in \mathbb{R}^{(d_u+d_v) \times |V|}$ are matrices, $\textbf{b}^1$ is $d_v$, and $\textbf{b}^2$ is $|V|$ dimensional vectors.

\end{document}